\documentclass[journal,twocolumn,10pt,twoside]{IEEEtran}
\usepackage{amsmath,amsopn,amssymb,bm}
\usepackage{amsfonts}
\usepackage{gensymb}
\usepackage{amsthm} 
\usepackage{graphicx,xspace,color}
\usepackage{epsfig}
\usepackage{longtable,multirow}
\usepackage{array}
\usepackage{stfloats}
\usepackage{cuted}
\usepackage{cite}
\usepackage[nolist]{acronym}
\usepackage{soul}
\usepackage{algorithm}
\usepackage[noend]{algpseudocode}
\usepackage{enumerate}
\usepackage{lettrine}
\usepackage{mathtools}
\usepackage{comment}
\usepackage{subcaption}
\usepackage{setspace} 
\usepackage[font=small]{caption}
\usepackage{booktabs}
\usepackage{booktabs}
\graphicspath{ {./figures/},{./photos/}}

\newtheorem{remark}{Remark}

\newtheorem{proposition}{Proposition}

\begin{document}

\begin{acronym}
	\acro{mmWave}{millimeter wave}
	\acro{THz}{Terahertz}
	\acro{EE}{energy efficiency}
	\acro{BS}{base station}
	\acro{Rx}{receiver}
	\acro{ULA}{uniform linear array}
	\acro{UPA}{uniform planar array}
	\acro{NULA}{nonuniform linear array}
	\acro{NUPA}{nonuniform planar array}
	\acro{AoSA}{arrays-of-subarrays}
	\acro{OMP}{orthogonal matching pursuit}
	\acro{SNR}{signal-to-noise ratio}
	\acro{SINR}{signal-to-interference-plus-noise ratio}
	\acro{CDF}{cumulative distribution function}
	\acro{MRT}{maximum ratio transmission}
	\acro{ZF}{zero forcing}
	\acro{LoS}{line-of-sight}
	\acro{NLoS}{non-line-of-sight}
	\acro{i.i.d.}{independent and identically}
	\acro{RF}{radio frequency}
	\acro{MIMO}{multiple-input multiple-output}
	\acro{CS}{compressed sensing}
	\acro{AoA}{angles-of-arrival}
	\acro{TDD}{time-division duplex}
	\acro{DAC}{digital-to-analog converter}
	\acro{NSE}{normalized squared error}
	\acro{CSI}{channel state information}
\end{acronym}

\title{Superdirective Antenna Pairs for Energy-Efficient Terahertz Massive MIMO}

\author{Konstantinos Dovelos, Stylianos D. Assimonis, Hien Quoc Ngo,~\IEEEmembership{Senior Member,~IEEE}, and Michail Matthaiou,~\IEEEmembership{Fellow,~IEEE}

\thanks{Manuscript received July 1, 2022; revised December 19, 2022, May 6, 2023, and August 8, 2023; accepted August 22, 2023.}

\thanks{Konstantinos Dovelos is with Meta Materials. Inc., Athens, Greece, email: kostis.dovelos@metamaterial.com. All other authors are with the Centre for Wireless Innovation, Queen's University Belfast, Belfast BT3 9DT, U.K., email: \{s.assimonis, hien.ngo, m.matthaiou\}@qub.ac.uk.}
}
\maketitle

\begin{abstract}
\ac{THz} communication is widely deemed the next frontier of wireless networks owing to the abundant spectrum resources in the \ac{THz} band. Whilst \ac{THz} signals suffer from severe propagation losses, a massive antenna array can be deployed at the \ac{BS} to mitigate those losses through beamforming. Nevertheless, a very large number of antennas increases the \ac{BS}'s hardware complexity and power consumption, and hence it can lead to poor \ac{EE}. To surmount this fundamental problem, we propose a novel array design based on superdirectivity and nonuniform inter-element spacing. Specifically, we exploit the mutual coupling between closely spaced elements to form superdirective pairs. A unique property of them is that all require the same excitation amplitude, and thus can be driven by a single radio frequency chain akin to conventional phased arrays. Moreover, they facilitate multi-port impedance matching, which ensures maximum power transfer for any beamforming angle. After addressing the implementation issues of superdirectivity, we show that the number of BS antennas can be effectively reduced without sacrificing the achievable rate. Simulation results demonstrate that our design offers huge \ac{EE} gains compared to uncoupled arrays with uniform spacing, and hence could be a radical solution for future \ac{THz} systems.
\end{abstract}


\section{Introduction}
Massive \ac{MIMO} is now a mature technology, which has been adopted by 5G new radio to provide superior network capacity and coverage. In parallel, \ac{mmWave} systems start gaining ground as an effective way for delivering multi-gigabit rates thanks to their very large   bandwidths~\cite{mmwave_5g,mmwave_oam_array}. Toward this direction, communication above $100$ GHz, i.e., terahertz (THz) frequencies, is widely deemed the next frontier of wireless systems with a plethora of promising applications, ranging from ultra-broadband femtocells to terabit-per-second links for wireless backhaul~\cite{6g_and_beyond}. Nevertheless, \ac{THz} signals are subject to severe propagation and molecular absorption losses, which can drastically limit the communication range and coverage~\cite{the_road_to_6g}. To deal with this problem, 
large antenna arrays can be deployed at the \acf{BS} to increase the signal power by means of sharp beamforming. As a result, massive \ac{MIMO} is expected to be an integral component of future \ac{THz} systems~\cite{prospective_antenna_tech}. On the other hand, \ac{THz} \ac{RF} circuits, e.g., power amplifiers, phase shifters, etc., exhibit significantly higher power consumption than their sub-6~GHz counterparts~\cite{thz_challenges}. Additionally, baseband processing with multiple \ac{RF} chains during channel estimation and data transmission is power intensive~\cite{mMIMO_theanswer}. In conclusion, achieving a large beamforming gain in an energy efficient manner constitutes a major engineering challenge which calls for novel solutions.

The beamforming capabilities of multi-antenna systems have been extensively investigated in the past. For example, a typical $N$-element phased array offers a gain that scales linearly with $N$ when \ac{MRT} is used~\cite{beam_steering_mmwave_array}. However, higher gains are possible by leveraging the mutual coupling between adjacent antennas. Specifically, Uzkov theoretically proved in his seminal work~\cite{uzkov_paper} that a \ac{ULA} of $N$ isotropic radiators and vanishingly small inter-element spacing has an endfire directivity of $N^2$, a phenomenon now known as \textit{superdirectivity}. 

The theme of superdirectivity has been widely studied from both information theoretic and pure electromagnetic standpoints, e.g.,~\cite{marzetta_sd1,marzetta_sd2, sp_mimo_1, sp_mimo_2,com_model_lis,sp_dof,small_parasitic_sdarray, sd_linear_dipoles_opt}, and references therein. However, most of the related literature considers arrays of uniform spacing. More importantly, it overlooks various communication and practical aspects, such as the \ac{EE}, implementation limitations in hybrid analog-digital architectures, and channel estimation performance. Consequently, there are still critical questions about how \ac{THz} massive \ac{MIMO} could fully benefit from superdirectivity. Regarding other approaches, there is a stream of recent papers on \ac{AoSA} with metallic antennas~\cite{aosa_thz,dynamic_aosa_mmwave,dynamic_aosa_thz} and graphene-based plasmonic nanoantennas~\cite{graphene_nanoantennas1, graphene_nanoantennas2}, as well as on intelligent reflecting surfaces~\cite{icc_2021, thz_irs_2}, which can improve the~\ac{EE} of the system. Yet, they neglect the superdirective effects of closely spaced \ac{BS} antennas.

This paper aims to show that superdirectivity can be ingeniously used to reduce the hardware complexity and boost the~\ac{EE} of \ac{THz} massive \ac{MIMO}. The contributions of our work are summarized as follows:
\begin{itemize}
\item We introduce a coupling-aware array model based on \textit{antenna theory}. In particular, we derive the input impedance matrix of the \ac{BS} array assuming lossy dipole antennas. Note that directional antennas, such as linear dipoles, are mutually coupled even for half-wavelength spacing. Therefore, proper characterization of their electromagnetic interaction is crucial to beamforming~\cite{opt_bf_mc}.
\item Based on the introduced array model, we study the implementation issues of superdirectivity. In particular, we look into the impedance matching problem as well as the realization of superdirective beamsteering in a hybrid analog-digital array architecture. We address both problems by proposing a novel array design relying on coupled antenna pairs. Specifically, the \ac{BS} array is divided into multiple two-element groups, which are adequately separated so that inter-group coupling can be neglected. The resulting \ac{NULA} greatly simplifies the optimal multi-port matching, and more importantly, requires uniform amplitude excitation. Consequently, it can be driven by a single \ac{RF} chain to produce a pencil-like beam, similar to conventional phased arrays. The presented structure is also extended to the \ac{NUPA} case. It is worth stressing that our method can be readily applied to an \ac{AoSA}, wherein each subarray is a \ac{NULA} or a \ac{NUPA}.
\item We exploit the excessive power gain of the proposed design to decrease the number of \ac{BS} antennas. This leads to a low-dimensional massive \ac{MIMO} system, whose performance is assessed in terms of the achievable rate under perfect and imperfect \ac{CSI}. For this purpose, approximate closed-form expressions for the signal and interference powers are provided assuming~\ac{MRT}. Moreover, the channel estimation problem is addressed by leveraging the popular \ac{OMP} algorithm.
\item Extensive simulation results are provided corroborating our analysis. In particular, it is demonstrated that the proposed array design boosts the \ac{EE} of \ac{THz} massive \ac{MIMO} without sacrificing the data transmission and channel estimation performances. As such, it has the potential to realize low-complexity and energy-efficient \ac{MIMO} arrays with sharp beamforming capabilities.
\end{itemize}
 \begin{figure}[t]
 	\centering
 	\includegraphics[width=0.74\linewidth]{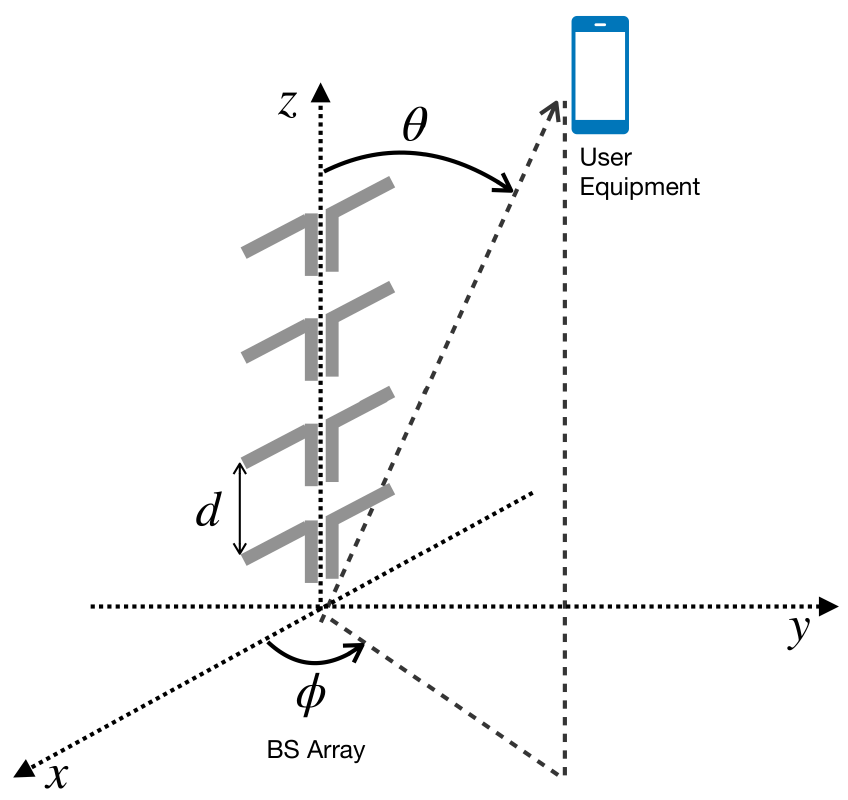}
 	\caption{Illustration of the antenna array considered at the \ac{BS}.}
 	\label{fig:Fig1}
 \end{figure}
 \begin{table}[t]
 	\centering
 	\caption{Main Notation used in this Work}
 	\label{Table:Notation}
 	\small
 	\begin{tabular}{|l  l|}
 		\hline
 		\textbf{Notation} & \textbf{Description} \\
 		\hline
 		$N$ & Number of BS antennas\\
 		$d$ & Inter-element spacing in \ac{ULA} \\
 		$N_g$ & Number of antenna groups in \ac{NULA} \\
 		$\bar{N}$ & Number of antennas per group in \ac{NULA}\\
 		$d_g$ & Inter-group spacing \\
 		$\bar{d}$ & Inter-element spacing within each antenna group \\
 		$f$, $\lambda, \kappa$ & Carrier frequency, wavelength, wavenumber \\
 		$\ell$, $\rho$ & Dipole length, radius \\
 		$\eta, \mu$ & Free-space impedance, permeability \\
 		$\sigma_c$ & Copper conductivity\\
 		$\mathbf{Z}_{\text{ideal}}$ & Input impedance matrix for lossless dipoles\\
 		$\mathbf{Z}$ & Input impedance matrix for lossy dipoles\\
 		\hline
 	\end{tabular}
 \end{table}
The rest of the paper is organized as follows: Section~\ref{sec:array_model} delineates the \ac{BS} array model. Section~\ref{sec:impedance_matching} presents the impedance matching problem. Section~\ref{sec:implementation_sd} delves into the implementation issues of superdirectivity and details the proposed solution. Section~\ref{sec:system_model} introduces the system model used for performance evaluation. Section~\ref{sec:si_powers} analyzes the signal and interference powers under the proposed array. Section~\ref{sec:channel_est} addresses the channel estimation problem. Section~\ref{sec:reduced_num_antennas} explains how to reduce the number of \ac{BS} antennas. Section~\ref{sec:numerical_results} is devoted to numerical simulations. Finally, Section~\ref{sec:conclusions} summarizes the main conclusions of this~work.

\textit{Notation}: Throughout the paper, $D_{N}(x) = \frac{\sin(Nx/2)}{N\sin(x/2)}$ is the Dirichlet sinc function; $\mathbf{A}(\cdot,\cdot,\cdot)$ is a vector field; $\mathbf{A}$ is a matrix; $\mathbf{A}^{*}$, $\mathbf{A}^{\dagger}$, $\mathbf{A}^{H}$, and $\mathbf{A}^{T}$ are the conjugate, pseudoinverse, conjugate transpose, and transpose of $\mathbf{A}$, respectively; $[\mathbf{A}]_{i,j}$ is the $(i,j)$th entry of $\mathbf{A}$; $\mathbf{A}(i)$ is the $i$th column of $\mathbf{A}$; $\text{blkdiag}(\mathbf{A}_1,\dots, \mathbf{A}_n)$ is a block diagonal matrix; $\mathbf{a}$ is a vector; $\|\mathbf{a}\|_1$ and $\|\mathbf{a}\|$ are the $l_1$-norm and $l_2$-norm of~$\mathbf{a}$, respectively; $\text{mag}(\mathbf{a}) = [|a_1|,\dots, |a_N|]^T$ for $\mathbf{a}= [a_1,\dots, a_N]^T$; $\mathbf{1}_{N\times M}$ is the $N\times M$ matrix with unit entries; $\mathbf{I}_{N}$ is the $N\times N$ identity matrix; $\mathbf{0}_{N\times M}$ is the $N\times M$ matrix with zero entries; $\otimes$ denotes the Kronecker product; $\mathbf{a}\cdot\mathbf{b}$ is the inner product between $\mathbf{a}$ and $\mathbf{b}$; $\mathbb{E}\{\cdot\}$ denotes expectation; $\mathcal{CN}(\bm{\mu}, \mathbf{R})$ is a complex Gaussian vector with mean $\bm{\mu}$ and covariance matrix~$\mathbf{R}$. Finally, $\text{Re}\{\cdot\}$ and $\text{Im}\{\cdot\}$ are the real and imaginary parts of a complex variable, respectively.

\section{Antenna Array Model}\label{sec:array_model}
In this section, we present the \ac{BS} array model which takes into account antenna mutual coupling.  

\subsection{Radiated Power}
Consider an $N$-element \ac{ULA} along the $z$-axis at the \ac{BS}, as depicted in Fig.~\ref{fig:Fig1}.\footnote{We consider a linear array at the \ac{BS} since it constitutes the building block of planar arrays. The planar case is investigated in Section~\ref{sec:planar_case} and thereafter.} The inter-element spacing is $d$. Each element is a linear dipole parallel to the $x$-axis, and is of length~$\ell$ and radius~$\rho$. According to~\cite[Ch. 4]{balanis_book}, the current distribution on each dipole $n$ has approximately the sinusoidal~form
\begin{equation}\label{eq:sin_current_approximation}
I_n(x')  \approx I_n(0)\frac{\sin\left(\kappa\ell/2-\kappa|x'|\right)}{\sin(\kappa\ell/2)}, \quad |x'| \leq \ell/2,
\end{equation}
where $I_n(0)\in\mathbb{C}$ is the input current, $\kappa = 2\pi/\lambda$ is the wavenumber, and $\lambda$ is the wavelength. We next focus on an arbitrary user who is in the far field of the \ac{BS} array. The user's location is described by the tuple $(r\cos\phi\sin\theta,r\sin\phi\sin\theta,r\cos\theta)$, where $r$, $\theta\in[0,\pi]$, and $\phi\in[0,2\pi]$ are the radial distance, polar angle, and azimuth angle, respectively. The electric field at the user is then specified as (see Appendix A)
\begin{align}\label{eq:e_field}
\mathbf{E}(r,\theta,\phi) &=  -j\eta\frac{e^{-j \kappa r}}{2\pi r }\sum_{n=0}^{N-1} e^{j\kappa \hat{\mathbf{r}}\cdot \mathbf{r}_n} I_n(0)\mathbf{F}(\theta,\phi),
\end{align}
where
\begin{align}\
\mathbf{F}(\theta,\phi) &= \frac{\cos(\kappa\ell/2\cos\phi\sin\theta) - \cos(\kappa\ell/2)}{\sin(\kappa\ell/2)(\sin^2\phi + \cos^2\phi\cos^2\theta)} \nonumber \\
&\times 
( \cos\theta\cos\phi \mathbf{e}_{\theta} - \sin\phi \mathbf{e}_{\phi}), 
\end{align}
is the vector field pattern of each dipole, $\mathbf{e}_{\theta}$ and $\mathbf{e}_{\phi}$ are the unit vectors along the polar and azimuth directions, respectively, $\eta$ is the characteristic impedance of free-space, $\hat{\mathbf{r}} = (\cos\phi\sin\theta,\sin\phi\sin\theta,\cos\theta)^T$ is the unit radial vector along the user direction, and $\mathbf{r}_n = (0,0,nd)$ is the position vector of the $n$th antenna. The radiation intensity [W/sr] is written in vector form as
\begin{equation}\label{radiation_intensity}
U \triangleq \frac{\|\mathbf{E}(r,\theta,\phi)\|^2}{2\eta} r^2= \frac{\eta}{8\pi^2}\|\mathbf{F}(\theta,\phi)\|^2\left|\mathbf{a}^H(\theta)\mathbf{i}\right|^2,
\end{equation}
where $\mathbf{a}(\theta) = [e^{-j\kappa \hat{\mathbf{r}}\cdot \mathbf{r}_0},\dots,e^{-j\kappa \hat{\mathbf{r}}\cdot \mathbf{r}_{N-1}}]^T\in\mathbb{C}^{N\times 1}$ and $\mathbf{i} = [I_0(0),\dots, I_{N-1}(0)]^T\in \mathbb{C}^{N\times 1}$ are the far-field array response vector and the vector of input currents, respectively. Using~\eqref{radiation_intensity}, the power radiated by the antenna array is 
\begin{align}\label{radiated_power_em}
&P_{\text{rad}} = \int_0^{2\pi}\!  \int_0^{\pi} U \sin\theta \text{d}\theta \text{d}\phi  
\nonumber \\
&= \frac{1}{2}\mathbf{i}^H\underbrace{\left( \frac{\eta }{4\pi^2}\!\!\int_0^{2\pi}  \int_0^{\pi} \mathbf{a}(\theta)\mathbf{a}^H(\theta)\|\mathbf{F}(\theta,\phi)\|^2\sin\theta \text{d}\theta \text{d}\phi,\right)}_{\text{Re}\{\mathbf{Z}_{\text{ideal}}\} } \mathbf{i} \nonumber \\
& = \frac{1}{2}\mathbf{i}^H\text{Re}\{\mathbf{Z}_{\text{ideal}}\}\mathbf{i},
\end{align}
where $\mathbf{Z}_{\text{ideal}}\in\mathbb{C}^{N\times N}$ is the input impedance matrix of the array assuming \textit{lossless} antennas. Moreover, $R_i\triangleq\left[\text{Re}\{\mathbf{Z}_{\text{ideal}}\}\right]_{n,n}  =  \frac{\eta }{4\pi^2}\int_0^{2\pi}\!  \int_0^{\pi} \|\mathbf{F}(\theta,\phi)\|^2\sin\theta \text{d}\theta \text{d}\phi$ is the input resistance\footnote{Recall that the input resistance equals the radiation resistance divided by $\sin^2(\kappa\ell/2)$~\cite[Ch.~8]{balanis_book}. This is because all quantities are expressed in terms of the input currents rather than the current maxima $\{I_n(0)/\sin(\kappa\ell/2)\}_{n=0}^{N-1}$.} of each lossless dipole.

\subsection{Input Power and Array Gain}
Realistic dipole antennas exhibit a conduction/loss resistance which leads to heat dissipation. Because of the \textit{skin effect} of conductive wires carrying an alternating current, the loss resistance per unit length is given by~\cite[Eq. (2-90b)]{balanis_book}
\begin{equation}\label{eq:loss_resistance_pul}
\bar{R}_{\text{loss}} = \frac{1}{2\rho}\sqrt{\frac{\mu f}{\pi\sigma_c}},
\end{equation}
where $f$ is the carrier frequency, $\mu$ is the permeability of free-space, and $\sigma_c$ is the conductivity of the wire material. Under the sinusoidal current distribution in~\eqref{eq:sin_current_approximation}, the loss resistance relative to the input current $I_n(0)$ is then specified as
\begin{align}\label{eq:loss_resistance}
R_{\text{loss}} = \bar{R}_{\text{loss}}
\int_{-\ell/2}^{
	\ell/2} \left|\frac{I_n(x')}{I_n(0)}\right|^2 \text{d}x'  
=  \frac{\kappa  \ell - \sin(\kappa\ell)}{4 \kappa \rho\sin^2\left(\kappa\ell/2\right)}\sqrt{\frac{\mu f}{\pi\sigma_c}}, 
\end{align}
which yields the overall power loss~\cite{super_gain_array} 
\begin{equation}
P_{\text{loss}} = \frac{1}{2}\sum_{n=0}^{N-1}R_{\text{loss}}|I_n(0)|^2 = \frac{1}{2}R_{\text{loss}}\|\mathbf{i}\|^2.
\end{equation}
Consequently, the input power at the antenna ports is 
\begin{align}
P_{\text{in}} &= P_{\text{loss}} + P_{\text{rad}} \nonumber \\
&= \frac{1}{2}R_{\text{loss}}\|\mathbf{i}\|^2 +  \frac{1}{2}\mathbf{i}^H\text{Re}\{\mathbf{Z}_{\text{ideal}}\}\mathbf{i}= \frac{1}{2}\mathbf{i}^H\text{Re}\{\mathbf{Z}\} \mathbf{i},
\end{align}
where $\mathbf{Z} \triangleq R_{\text{loss}}\mathbf{I}_N + \mathbf{Z}_{\text{ideal}}$ is the impedance matrix of the lossy array. Finally, the array gain is defined~as
\begin{align}\label{eq:array_gain_general}
G(\theta,\phi)  & \triangleq \frac{4\pi U}{P_{\text{in}}}
= G_e(\theta,\phi)(R_{\text{loss}} + R_i)\frac{|\mathbf{a}^H(\theta)\mathbf{i}|^2}{\mathbf{i}^H\text{Re}\{\mathbf{Z}\}\mathbf{i}},
\end{align}
where $G_e(\theta,\phi) = \frac{\eta \|\mathbf{F}(\theta,\phi)\|^2}{\pi (R_{\text{loss}} + R_i)}$ denotes the gain of each dipole. 

\subsection{Optimal Currents under Fixed Input Power}
According to Friis transmission formula, the power received by the user is given by~\cite{balanis_book}
\begin{equation}
P_r = P_{\text{in}}\left(\frac{\lambda}{4\pi r}\right)^2G(\theta,\phi),
\end{equation}
where an isotropic antenna has been assumed at the user for simplicity. We now seek to find $\mathbf{i}$ that maximizes the received power (or equivalently $G(\theta,\phi)$) subject to the input power constraint $P_{\text{in}} \leq P_t$. The objective~\eqref{eq:array_gain_general} is a generalized Rayleigh quotient. Thus, the optimal current excitation is obtained~as~\cite{super_gain_array}
\begin{equation}\label{eq:opt_current}
\mathbf{i}^{\text{opt}}= \sqrt{\frac{2P_t}{\mathbf{a}^H(\theta)\text{Re}\{\mathbf{Z}\} ^{-1}\mathbf{a}(\theta)}}\text{Re}\{\mathbf{Z}\} ^{-1}\mathbf{a}(\theta),
\end{equation}
and the maximum array gain is
\begin{align}
G_{\max}(\theta,\phi) &= G_e(\theta,\phi)(R_{\text{loss}} + R_i) \mathbf{a}^H(\theta)\text{Re}\{\mathbf{Z}\}^{-1}\mathbf{a}(\theta).
\end{align}

\begin{remark}[Uncoupled ULA]
In the absence of mutual coupling, $\emph{Re}\{\mathbf{Z}_{\emph{ideal}}\} = R_i\mathbf{I}_N$ and $P_{\emph{rad}} = \frac{1}{2}R_i\|\mathbf{i}\|^2$. Moreover, $\mathbf{i}^{\emph{opt}} =\sqrt{\frac{2P_t}{N(R_{\emph{loss}} + R_i )}}\mathbf{a}(\theta)$ and $G_{\max}(\theta,\phi)= G_e(\theta,\phi) N$, which is the typical $O(N)$ power gain.
\end{remark}
\begin{figure}[t]
	\centering
	\includegraphics[width=0.95\linewidth]{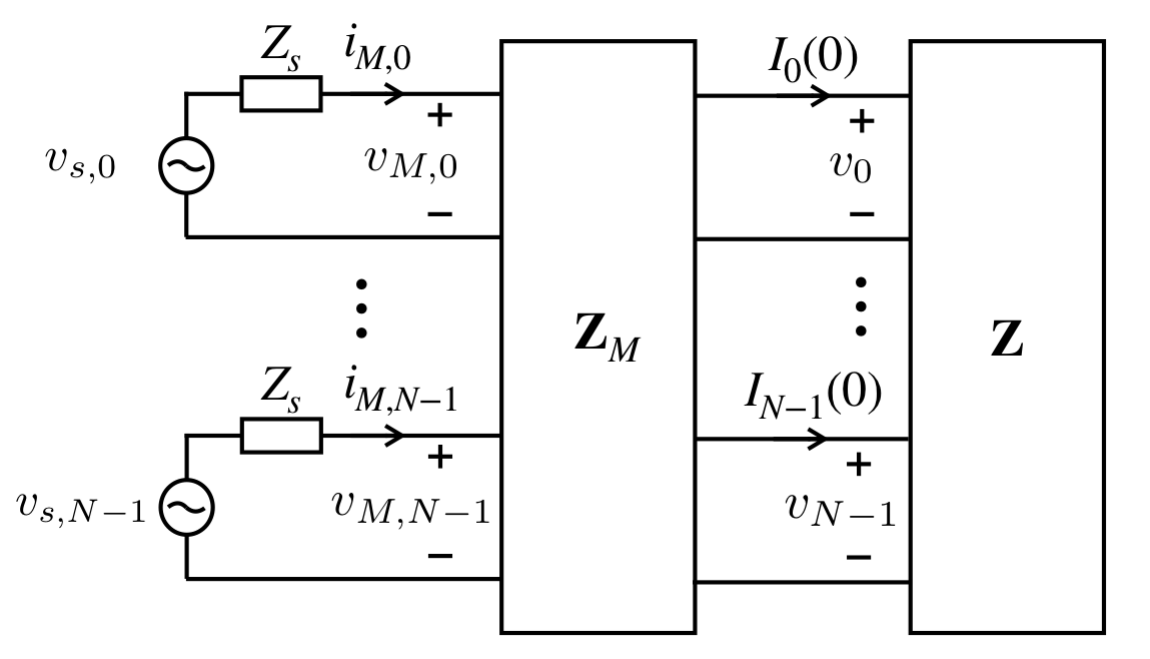}
	\caption{Equivalent multi-port network of an antenna array covering signal generation, impedance matching, and mutual coupling.}
	\label{fig:Fig2}
\end{figure}
\begin{figure}[t]
	\centering
	\includegraphics[width=1\linewidth]{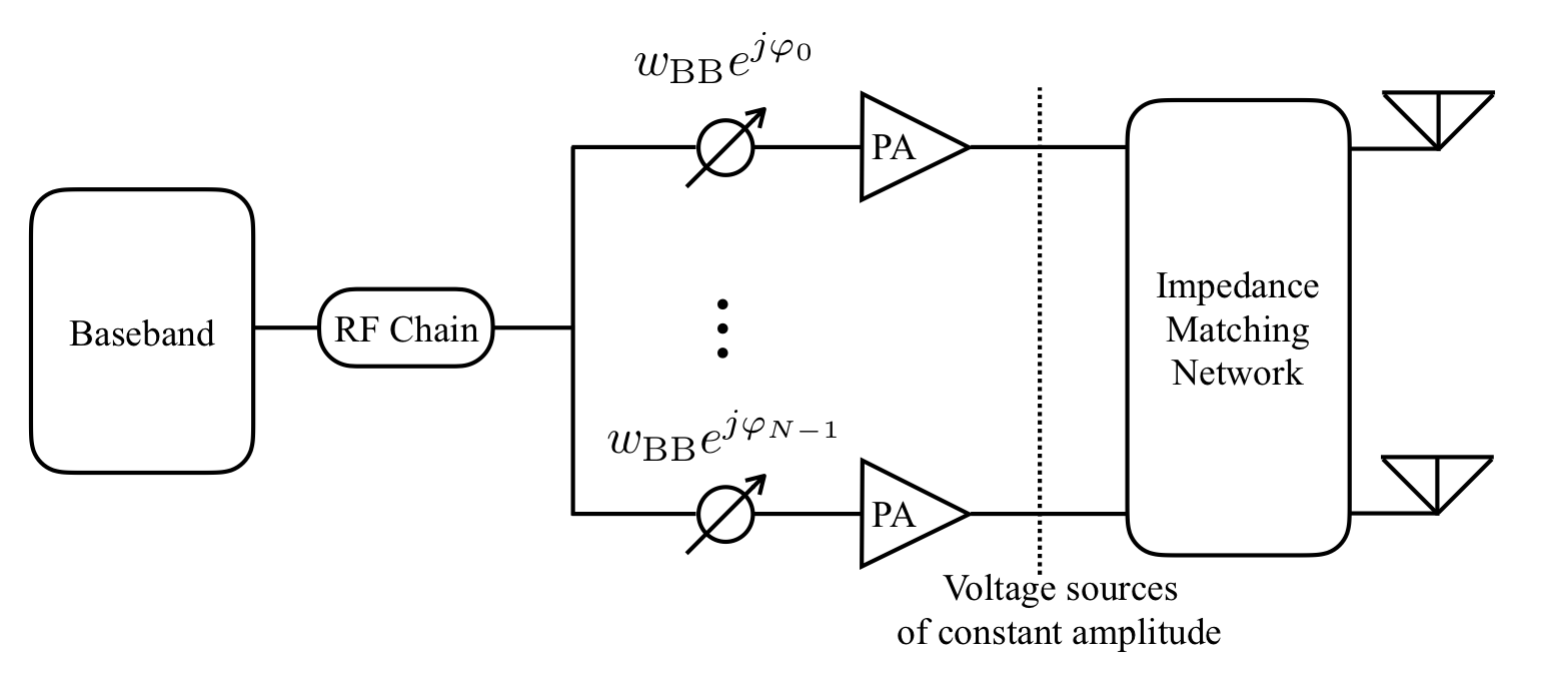}
	\caption{Block diagram of hybrid array architecture with one RF chain.}
	\label{fig:Fig21}
\end{figure}
\section{Impedance Matching in the Presence of Mutual Coupling}\label{sec:impedance_matching}

Mutual coupling alters the input impedance of each dipole. To see this, let  $\mathbf{v}=[v_{0},\dots,v_{N-1}]^T\in\mathbb{C}^{N\times 1}$ denote the vector of voltages at the antenna ports. Then, we have that $\mathbf{v}=  \mathbf{Z}\mathbf{i} = \mathbf{Z}_a\mathbf{i}$, where $\mathbf{Z}_a\in\mathbb{C}^{N\times N}$ is a diagonal matrix with entries~\cite[Ch. 8]{balanis_book}
\begin{equation}\label{eq:active_impedance}
[\mathbf{Z}_a]_{n,n} \triangleq \left([\mathbf{Z}]_{n,n} + \sum_{m=0,m\neq n}^{N-1}[\mathbf{Z}]_{n,m}\frac{I_m(0)}{I_n(0)} \right),
\end{equation}
where $[\mathbf{Z}_a]_{n,n}$ is the \textit{active} impedance of the $n$th antenna. The active impedance of each  dipole hinges on the excitation currents, and hence single-port conjugate matching is optimal only for a specific beamforming angle $\theta$~\cite{opt_single_port1}. For this reason, we resort to multi-port matching and model the \ac{BS} array as in~Fig.~\ref{fig:Fig2}. The impedance network is a passive and \textit{lossless} $2N$-port network described by the matrix $\mathbf{Z}_M\in\mathbb{C}^{2N\times 2N}$, which is partitioned as
\begin{equation}
\mathbf{Z}_M =   
\begin{bmatrix}
\mathbf{Z}_{M11} & \mathbf{Z}_{M12} \\ 
\mathbf{Z}_{M21} & \mathbf{Z}_{M22}
\end{bmatrix}.
\end{equation}
The lossless property implies that $\mathbf{Z}_M$ has only imaginary entries. The vector of voltage sources is denoted by $\mathbf{v}_s=[v_{s,0},\dots,v_{s,N-1}]^T\in\mathbb{C}^{N\times 1}$. Each voltage source has an internal impedance $Z_s\in\mathbb{C}$, with $\text{Re}\{Z_s\} = R_s$. Likewise, the voltages and currents at the input ports of the network are denoted by $\mathbf{v}_M=[v_{M,0},\dots,v_{M,N-1}]^T\in\mathbb{C}^{N\times 1}$ and $\mathbf{i}_M=[i_{M,0},\dots,i_{M,N-1}]^T\in\mathbb{C}^{N\times 1}$, respectively. Based on basic circuit analysis, the relationship between the voltages and currents at the input and output of the matching network~is~\cite{circuit_th_commun} 
\begin{equation}\label{eq:matching_network}
\begin{bmatrix}
\mathbf{v}_M \\ 
\mathbf{v}
\end{bmatrix} = \begin{bmatrix}
\mathbf{Z}_{M11} & \mathbf{Z}_{M12} \\ 
\mathbf{Z}_{M21} & \mathbf{Z}_{M22}
\end{bmatrix}\begin{bmatrix}
\mathbf{i}_M \\ 
-\mathbf{i}
\end{bmatrix}.
\end{equation}
Using~\eqref{eq:matching_network} and the relationship $\mathbf{v}=\mathbf{Z}\mathbf{i}$ yields 
\begin{equation}
\mathbf{v}_M = \underbrace{\left(\mathbf{Z}_{M11} - \mathbf{Z}_{M12}(\mathbf{Z} +  \mathbf{Z}_{M22})^{-1}\mathbf{Z}_{M21} \right)}_{\mathbf{Z}_T}\mathbf{i}_M,
\end{equation}
where $\mathbf{Z}_T\in\mathbb{C}^{N\times N}$ is the overall transmit impedance matrix accounting for power matching and mutual coupling. The problem of optimal multi-port matching is to find $\mathbf{Z}_M$ so that $\mathbf{Z}_T = Z^*_s\mathbf{I}_N$. In this case, there are no reflection losses and half of the generated power enters the antenna array, i.e., maximum power transfer~\cite{pozar_book}. This is accomplished for~\cite{circuit_th_commun}
\begin{equation}\label{eq:optimal_matching}
\mathbf{Z}_M =  \begin{bmatrix}
-j\text{Im}\{Z_s\} \mathbf{I}_N  & -j\sqrt{R_s}\text{Re}\{\mathbf{Z}\}^{1/2} \\ 
-j\sqrt{R_s}\text{Re}\{\mathbf{Z}\}^{1/2} & -j\text{Im}\{\mathbf{Z}\}
\end{bmatrix}.
\end{equation} 
Under \eqref{eq:optimal_matching}, we have that
\begin{align}\label{eq:source_voltages}
\mathbf{v}_s = Z_s\mathbf{I}_N\mathbf{i}_M + \mathbf{v}_M 
= \left(Z_s\mathbf{I}_N + \mathbf{Z}_T\right)\mathbf{i}_M = 2R_s\mathbf{i}_M,
\end{align}
and the total power generated by the voltage sources is
\begin{equation}
P_{\text{total}} =  \frac{1}{2}\text{Re}\left\{\mathbf{v}^H_s\mathbf{i}_M\right\} = R_s\|\mathbf{i}_M\|^2.
\end{equation}	
From~\eqref{eq:matching_network} and~\eqref{eq:optimal_matching}, it also holds that
\begin{align}\label{eq:i_M}
		\mathbf{i}_M &=  \mathbf{Z}_{M21}^{-1}(\mathbf{Z} +  \mathbf{Z}_{M22})\mathbf{i} \nonumber \\
		&= \frac{j}{\sqrt{R_s}}\text{Re}\{\mathbf{Z}\}^{1/2}\mathbf{i},
\end{align}
and hence $P_{\text{total}} = \mathbf{i}^H\text{Re}\{\mathbf{Z}\}\mathbf{i} = 2 P_{\text{in}}$, which confirms the optimality of~\eqref{eq:optimal_matching}.

\section{Hardware-Efficient Implementation of Superdirective Beamsteering}\label{sec:implementation_sd}
In this section, we investigate the problem of generating a superdirective beam with a single \ac{RF} chain and a low-complexity impedance matching network. 

\subsection{Problem Statement}
From~\eqref{eq:opt_current},~\eqref{eq:source_voltages}, and \eqref{eq:i_M}, the vector of voltage sources maximizing the array gain is given by
\begin{align}\label{eq:v_g_i}
\mathbf{v}^{\text{opt}}_s = 2R_s\mathbf{i}^{\text{opt}}_M &= j2\sqrt{R_s}\text{Re}\{\mathbf{Z}\}^{1/2}\mathbf{i}^{\text{opt}} \nonumber \\
&= j2\sqrt{\frac{2R_sP_t}{\mathbf{a}^H(\theta)\text{Re}\{\mathbf{Z}\} ^{-1}\mathbf{a}(\theta)}}\text{Re}\{\mathbf{Z}\}^{-1/2}\mathbf{a}(\theta).
\end{align}
In most massive \ac{MIMO} studies, beamforming is conveniently described by a complex vector $\mathbf{w}\in\mathbb{C}^{N\times 1}$ whose squared norm, $\|\mathbf{w}\|^2$, defines the transmit power. This representation originates from information theory and can be mapped to the voltage sources driving the array through the relationship~\cite{circuit_th_commun2}
\begin{align}
\mathbf{w} &= -\frac{1}{j 2\sqrt{R_s}}\left(\mathbf{v}_s^{\text{opt}}\right)^* \nonumber \\
 &= \sqrt{\frac{2 P_t}{\mathbf{a}^H(\theta)\text{Re}\{\mathbf{Z}\} ^{-1}\mathbf{a}(\theta)}}\text{Re}\{\mathbf{Z}\}^{-1/2}\mathbf{a}^*(\theta).
\end{align}
We next consider a hybrid analog-digital array with one \ac{RF} chain and $N$ analog phase shifters, as shown in Fig.~\ref{fig:Fig21}. In this case, $\mathbf{w} = w_{\text{BB}} [e^{j\varphi_0}, \dots,e^{j\varphi_{N-1}}]^T$, where $w_{\text{BB}}$ denotes the complex amplitude of baseband processing whilst $\{\varphi_n\}_{n=0}^{N-1}$ are the variable phases. Consequently, beamsteering requires voltage sources $\{v_{s,n}\}_{n=0}^{N-1}$ of uniform magnitude. From~\eqref{eq:v_g_i}, it is easy to verify that $\mathbf{v}^{\text{opt}}_s$ has entries of different magnitudes, and hence cannot be realized by a single \ac{RF} chain. In summary, superdirective \ac{ULA}s require amplitude control at each antenna, which can be provided by a fully digital array or an active phased array with digitally controlled amplitude attenuators. However, those architectures increase substantially the hardware complexity and power consumption. Similarly, multi-port matching requires interconnections and circuit components between all the $2N$ ports~\cite{complex_multiport_matching}. Thus, it is prohibitively complex for a large number $N$ of antennas~\cite{limits_mMIMO}.
 \begin{figure}[t]
 	\centering
 	\includegraphics[width=0.96\linewidth]{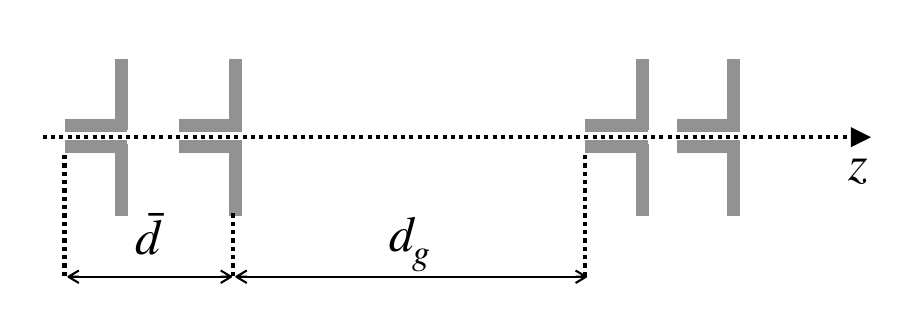}
 	\caption{Example of a \ac{NULA} with $N_g=\bar{N}=2$.}
 	\label{fig:Fig22}
 \end{figure}
\subsection{Proposed Solution}\label{sec:proposed_array}

\subsubsection{Nonuniform Linear Array}
The \ac{BS} array is divided into $N_g$ groups of $\bar{N}$ antennas each, i.e., $N=N_g\bar{N}$. Let $d_g$ and $\bar{d}$ denote the inter-group and inter-element spacings, respectively, as shown in Fig.~\ref{fig:Fig22}. The distance of the $\bar{n}$th element in the $n_g$th subarray from the origin of the coordinate system is given by $\left(\bar{n}\bar{d} + n_g(d_g + (\bar{N}-1)\bar{d})\right)\cos\theta$, where $\bar{n}=0,\dots, \bar{N}-1$ and $n_g=0,\dots, N_g-1$. Thus, the array response vector $\mathbf{a}(\theta)$ can be recast as
\begin{align}
\mathbf{a}(\theta) &= \left[\mathbf{a}_0(\theta),\mathbf{a}_0(\theta)e^{-j\kappa((d_g + (\bar{N}-1)\bar{d})\cos\theta)}, \dots,  \right. \nonumber\\
&\left.	 \quad \quad \quad \quad \mathbf{a}_0(\theta)e^{-j\kappa(N_g-1)(d_g + (\bar{N}-1)\bar{d})\cos\theta}\right]^T \nonumber\\
 &= \mathbf{a}_g(\theta)\otimes \mathbf{a}_0(\theta),
\end{align} 
where $\mathbf{a}_0(\theta) \triangleq [1,\dots, e^{-j\kappa(\bar{N}-1)\bar{d}\cos\theta}]^T\in\mathbb{C}^{\bar{N}\times 1}$ is the response vector of the $0$th antenna group, and $\mathbf{a}_g(\theta) \triangleq [1,\dots, e^{-j\kappa(N_g-1)(d_g + (\bar{N}-1)\bar{d})\cos\theta}]^T\in\mathbb{C}^{N_g\times 1}$ is the response vector between groups. Similarly, the input impedance matrix is partitioned as
\begin{equation}
\mathbf{Z} =  \begin{bmatrix}
\mathbf{Z}_{0}  & \mathbf{Z}_{0,1}   &\cdots &\mathbf{Z}_{0,N_g-1} \\
\mathbf{Z}_{1,0}  &\mathbf{Z}_{0}    &\cdots &\mathbf{Z}_{1,N_g-1}   \\
\vdots & \vdots  &\ddots & \vdots \\
\mathbf{Z}_{N_g-1,0}  &\mathbf{Z}_{N_g-1,1}   &\cdots & \mathbf{Z}_{0} 
\end{bmatrix}, 
\end{equation}
where $\text{Re}\{\mathbf{Z}_{n_{g1},n_{g2}}\}\in\mathbb{R}^{\bar{N}\times \bar{N}}$ and $\text{Re}\{\mathbf{Z}_0\}\in\mathbb{R}^{\bar{N}\times \bar{N}}$ are defined by \eqref{eq:impedance_mtx_nula1} and \eqref{eq:impedance_mtx_nula2} at the bottom of this page, respectively.
\begin{figure*}[b]
\hrulefill
\begin{align}\label{eq:impedance_mtx_nula1}
	\text{Re}\{\mathbf{Z}_{n_{g1},n_{g2}}\} &\triangleq  \frac{\eta }{4\pi^2}\int_0^{2\pi}\int_0^{\pi} \mathbf{a}_0(\theta)\mathbf{a}^H_0(\theta) e^{-j\kappa(n_{g1}-n_{g2})(d_g + (\bar{N}-1)\bar{d})\cos\theta}\|\mathbf{F}(\theta,\phi)\|^2\sin\theta \text{d}\theta \text{d}\phi,  \quad \text{for} \quad n_{g1}  \neq  n_{g2}, \\[0.15cm]
	\text{Re}\{\mathbf{Z}_{0}\} & \triangleq \text{Re}\{\mathbf{Z}_{n_{g1},n_{g2}}\} =  \frac{\eta }{4\pi^2}\int_0^{2\pi}\int_0^{\pi} \mathbf{a}_0(\theta)\mathbf{a}^H_0(\theta) \|\mathbf{F}(\theta,\phi)\|^2\sin\theta \text{d}\theta \text{d}\phi 
	+ R_{\text{loss}} \mathbf{I}_{\bar{N}},  \quad \text{for} \quad n_{g1}  =  n_{g2}.
	\label{eq:impedance_mtx_nula2}
	\end{align}
\end{figure*}  
In fact, $\mathbf{Z}_0$ corresponds to the input impedance matrix of each $\bar{N}$-element group. For $\bar{N}=2$, we~have  
\begin{equation}
\text{Re}\{\mathbf{Z}_0\} \triangleq\begin{bmatrix}
R_{\text{self}}   & R_m \\
R_m & R_{\text{self}} 
\end{bmatrix},
\end{equation}
where $R_{\text{self}}  = R_{\text{loss}} + R_i$ denotes the self impedance for notational convenience, and $R_m$ is the real part of the mutual impedance between two adjacent dipoles.

When the inter-group spacing is sufficiently large, the coupling between groups can be neglected. Thus, $\text{Re}\{\mathbf{Z}_{n_{g1},n_{g2}}\}\approx \mathbf{0}_{\bar{N}\times \bar{N}}$ for $n_{g1}\neq n_{g2}$, which makes $\mathbf{Z}$ approximately a block diagonal matrix, namely $\text{Re}\{\mathbf{Z}\}\approx \mathbf{I}_{N_g}\otimes \text{Re}\{\mathbf{Z}_0\} = \text{Re}\{\mathbf{Z}_{\text{approx}}\}$. We finally stress that in addition to a large~$d_g$, the mutual coupling between antenna groups can be further reduced through various decoupling techniques, such as electromagnetic band gap structures placed between them~\cite{ebg_structures}.

\subsubsection{Coupled Antenna Pairs}\label{sec:superdirective_pairs}
A major drawback of beamsteering under mutual coupling is the requirement of amplitude control at each voltage source. However, in a \ac{NULA}, all antenna groups share the same voltage amplitudes. This is because from~\eqref{eq:v_g_i}
\begin{align}\label{eq:current_k}
\mathbf{v}^{\text{opt}}_s &= \bar{v}\text{Re}\{\mathbf{Z}\}^{-1/2}\mathbf{a}(\theta) \nonumber \\
& \approx \bar{v} \text{Re}\{\mathbf{Z}_{\text{approx}}\}^{-1/2}\mathbf{a}(\theta)\nonumber \\
&=  \bar{v} \left(\mathbf{I}_{N_g}\otimes \text{Re}\{\mathbf{Z}_0\}^{-1/2}\right)(\mathbf{a}_g(\theta)\otimes \mathbf{a}_0(\theta)) \nonumber \\
&=  \bar{v} \mathbf{a}_g(\theta) \otimes \left(\text{Re}\{\mathbf{Z}_0\}^{-1/2}\mathbf{a}_0(\theta)\right),
\end{align}
where $\bar{v}= j2\sqrt{2R_sP_t/\mathbf{a}^H(\theta)\text{Re}\{\mathbf{Z}\}^{-1}\mathbf{a}(\theta)}$, which gives
\begin{align}
\text{mag}\left(\mathbf{v}^{\text{opt}}_s \right) &= \bar{v}\begin{bmatrix}
\text{mag}\left(\text{Re}\{\mathbf{Z}_0\}^{-1/2}\mathbf{a}_0(\theta)\right) \\
\vdots \\
\text{mag}\left(\text{Re}\{\mathbf{Z}_0\}^{-1/2}\mathbf{a}_0(\theta)\right)
\end{bmatrix}.
\end{align}
By leveraging this unique property of NULAs, we further propose two-element groups. The following proposition justifies the choice of $\bar{N}=2$.

\begin{proposition}\label{prop1}
For $\bar{N}=2$, $\emph{mag}\left(\emph{Re}\{\mathbf{Z}_0\}^{-1/2}\mathbf{a}_0(\theta)\right)$ reduces to~\eqref{eq:mag_pair}  at the top of the next page.
 \begin{proof}
See Appendix B.	
\end{proof}
\end{proposition}
\begin{figure*}[t]
\begin{align}\label{eq:mag_pair}
\text{mag}\left(\text{Re}\{\mathbf{Z}_0\}^{-1/2}\mathbf{a}_0(\theta)\right)  &= 
\mathbf{1}_{2\times 1}\frac{1}{\sqrt{2R_{\text{self}} + 2\sqrt{R_{\text{self}}^2 - R_m^2}}}\nonumber \\[0.1cm]
&\times\sqrt{\left(\frac{R_{\text{self}}}{\sqrt{R_{\text{self}}^2 - R_m^2}} + 1\right)^2 + \frac{R_m^2}{R_{\text{self}}^2 - R_m^2} -2\left(\frac{R_{\text{self}}}{\sqrt{R_{\text{self}}^2 - R_m^2}} + 1\right)\frac{R_m}{\sqrt{R_{\text{self}}^2 - R_m^2}} \cos\left(\kappa\bar{d}\cos\theta\right)}.
\end{align}	
\hrulefill
\end{figure*}  
According to Proposition~\ref{prop1}, antenna pairs result in \textit{uniform} excitation amplitudes $\{|v_{s,n}|\}_{n=0}^{N-1}$, thereby enabling the realization of superdirective beamsteering with a single \ac{RF} chain and $N$ analog phase shifters. Additionally, each antenna pair is separately matched using the four-port network (see Appendix~B)
\begin{equation}\label{eq:z_m_0}
\mathbf{Z}_{M,0} = 
\begin{bmatrix}
-j\text{Im}\{Z_s\} \mathbf{I}_2 & -j\sqrt{R_s}\text{Re}\{\mathbf{Z}_{0}\}^{1/2} \\
		-j\sqrt{R_s}\text{Re}\{\mathbf{Z}_0\}^{1/2} & -j\text{Im}\{\mathbf{Z}_0\} 
\end{bmatrix}.
\end{equation} 
Therefore, the proposed solution also facilitates the implementation of optimal multi-port matching, which would not be feasible in a coupled \ac{ULA} with a large number of elements. Note that our solution relies on the symmetric structure of $\mathbf{Z}_0 \in \mathbb{C}^{2\times 2}$, which holds for radiators beyond linear dipoles. The main reason we chose dipoles is mathematical tractability; otherwise, we would heavily rely on full-wave simulations to assess the antenna array gain. The problem of finding the optimal antenna type to facilitate superdirectivity is beyond the scope of the current work, and is a promising avenue for future research.

\section{\ac{THz} Massive MIMO Model}\label{sec:system_model}
In this section, we introduce the high-level system model used to evaluate the performance of the proposed \ac{NULA} in terms of the achievable rate and \ac{EE}.

\subsection{Signal Model}
Consider a \ac{THz} massive MIMO system, where the \ac{BS} serves $K\ll N$ single-antenna users. Let $\mathbf{h}_k\in\mathbb{C}^{N\times 1}$ and $\beta_k$ denote the small-scale fading channel and path loss of user $k$, respectively. The downlink channel of user $k$ is then specified as $\sqrt{\beta}_k\mathbf{h}^T_k$. To facilitate hardware implementation, a fully connected analog-digital array with $N_{\text{RF}}$ \ac{RF} chains is considered at the \ac{BS}~\cite{mmwave_hardware_constrained}. Thus, the precoder is decomposed as $\mathbf{W} = \mathbf{W}_{\text{RF}}\mathbf{W}_{\text{BB}} = [\mathbf{w}_1,\dots,\mathbf{w}_{N_{\text{RF}}}]\in\mathbb{C}^{N\times N_{\text{RF}}}$, where $\mathbf{W}_{\text{RF}}\in\mathbb{C}^{N\times N_{\text{RF}}}$ is the RF beamformer realized by analog phase shifters, whereas  $\mathbf{W}_{\text{BB}}\in\mathbb{C}^{N_{\text{RF}}\times N_{\text{RF}}}$ is the baseband precoder. Without loss of generality, we assume $N_{\text{RF}} = K$ hereafter. Let $\mathbf{x} = [x_1,\dots,x_K]^T\sim\mathcal{CN}(\mathbf{0}_{K\times 1}, \mathbf{I}_K)$ be the vector of users' data symbols. The transmitted signal is then given by $\mathbf{W}\mathbf{x}\in\mathbb{C}^{N\times 1}$, and should satisfy the power constraint
\begin{align}
\mathbb{E}\left\{\|\mathbf{W}\mathbf{x}\|^2\right\} = \sum_{k=1}^K\|\mathbf{w}_{k}\|^2 \leq 2P_t,
\end{align}
where $2P_t$ is the \textit{total power} under perfect impedance matching. Given the above, the received baseband signal at the $k$th user is written as
\begin{equation}
y_k = \sqrt{\beta_k}\mathbf{h}_k^T\mathbf{w}_kx_k + \sqrt{\beta_k}\sum_{i=1, i\neq k}^K\mathbf{h}_k^T\mathbf{w}_ix_i + n_k,
\end{equation}
where $n_k\sim\mathcal{CN}(0,\sigma^2)$ is the additive noise. Finally, the \ac{SINR} of user $k$ is 
\begin{equation}\label{eq:sinr}
\text{SINR}_k = \frac{\beta_k\left|\mathbf{h}_k^T\mathbf{w}_k\right|^2}{\beta_k\sum_{i=1, i\neq k}^K\left|\mathbf{h}_k^T\mathbf{w}_i\right|^2 + B\sigma^2},
\end{equation}
where $B\sigma^2$ is the noise power over the transmit bandwidth $B$.

\subsection{Channel Model}\label{sec:antenna_theory_model}
Because of the severe path attenuation in the \ac{THz} band, multi-path scattering is very limited. We therefore assume \ac{LoS} links between the \ac{BS} and users, akin to~\cite{fast_tracking_thz_mimo}. In the presence of mutual coupling at the \ac{BS} array, the channel vector of user $k$ is expressed as~\cite[Eq. (105)]{circuit_th_commun}
\begin{equation}\label{eq:channel_vector}
\mathbf{h}_k \triangleq \text{Re}\left\{\mathbf{\bar{Z}}\right\}^{-1/2}\mathbf{a}(\theta_k),
\end{equation}
where $\mathbf{\bar{Z}} \triangleq \frac{1}{R_{\text{loss}} + R_i}\mathbf{Z}$ is the \textit{normalized} input impedance matrix of the \ac{BS} array.\footnote{The normalized input impedance matrix is used for notational convenience. In this way, the array gain is recast as $G(\theta,\phi) = G_e(\theta,\phi)\frac{|\mathbf{a}^H(\theta)\mathbf{i}|^2}{\mathbf{i}^H \text{Re}\left\{\mathbf{\bar{Z}}\right\}\mathbf{i}}$, and the path loss coefficient or channel vector will not include the term $R_{\text{loss}} + R_i$. For example, $\text{Re}\left\{\mathbf{\bar{Z}}\right\} = \mathbf{I}_N$ and $\mathbf{h}_k = \mathbf{a}(\theta_k)$ in the absence of coupling.} Finally, because the molecular absorption losses are no longer negligible at \ac{THz} frequencies, the path loss coefficient is calculated as~\cite{how_many_antennas_thz}
\begin{equation}\label{eq:path_loss}
\beta_k = G_e(\theta_k,\phi_k)\left(\frac{\lambda}{4\pi r_k}\right)^2e^{-\kappa_{\text{abs}}r_k},
\end{equation}
where $G_e(\theta_k,\phi_k)$ is the gain of each \ac{BS} antenna, $\lambda$ is the carrier wavelength, $r_k$ is the distance from the \ac{BS} to user~$k$, and $\kappa_{\text{abs}}$ is the molecular absorption coefficient determined by the composition of the propagation medium~\cite{thz_channel_model}.

\subsection{Power Consumption Model}
For a fully-connected array structure, the overall power consumption is given by~\cite{ee_mmwave_lowresdac}
\begin{align}
P_c =  P_{\text{BB}} + K P_{\text{RF}} + KN P_{\text{PS}} + NP_{\text{PA}} + P_t + P_{\text{CE}},
\end{align}
where $P_{\text{BB}} $, $P_{\text{RF}}$, $P_{\text{PS}}$, $P_{\text{PA}}$, and $P_{\text{CE}}$ denote the powers consumed by a baseband unit, an \ac{RF} chain, a phase shifter, a power amplifier, and the channel estimation process, respectively. Moreover, each \ac{RF} chain comprises a \ac{DAC}, a local oscillator, and a mixer, and hence $P_{\text{RF}} = P_{\text{DAC}} + P_{\text{LO}} + P_{\text{M}}$. We finally stress that the power consumption of splitters and combiners is negligible, and hence is ignored~\cite{ee_mmwave_lowresdac}. 

\section{Signal and Interference Powers}\label{sec:si_powers}

\subsection{Proposed \ac{NULA}}
We consider \ac{MRT} at the \ac{BS}. Under equal power allocation, we have that  
\begin{align}\label{eq:mrt_general}
\mathbf{w}_k &= \sqrt{\frac{2P_t}{K}}\frac{\mathbf{h}^*_k}{\|\mathbf{h}_k\|} \nonumber \\
&=\sqrt{\frac{2P_t}{K\mathbf{a}^H(\theta_k) \text{Re}\left\{\mathbf{\bar{Z}}\right\}^{-1}\mathbf{a}(\theta_k)}} \text{Re}\left\{\mathbf{\bar{Z}}\right\}^{-\frac{1}{2}}\mathbf{a}^*(\theta_k),
\end{align}
and the power of the desired signal, normalized by $K/(2P_t)$,~is 
\begin{equation}\label{eq:signal_power_coupling}
\frac{K}{2P_t}\left|\mathbf{h}_k^T\mathbf{w}_k\right|^2 =\mathbf{a}^H(\theta_k) \text{Re}\left\{\mathbf{\bar{Z}}\right\}^{-1}\mathbf{a}(\theta_k).
\end{equation}
Due to the block-diagonal structure of $\mathbf{\bar{Z}}$,~\eqref{eq:signal_power_coupling} simplifies to 
\begin{align}\label{eq:signal_power_nula}
&\frac{K}{2P_t}\left|\mathbf{h}_k^T\mathbf{w}_k\right|^2 \approx \mathbf{a}^H(\theta_k)\text{Re}\left\{\mathbf{\bar{Z}}_{\text{approx}}\right\}^{-1}\mathbf{a}(\theta_k) \nonumber  \\
&= \left(\mathbf{a}^H_g(\theta_k)\otimes \mathbf{a}^H_0(\theta_k)\right)\!\left(\mathbf{I}_{N_g}\otimes \text{Re}\left\{\bar{\mathbf{Z}}_0\right\}^{-1}\right)\!\left(\mathbf{a}_g(\theta_k)\otimes \mathbf{a}_0(\theta_k)\right) \nonumber \\
& = N_g \mathbf{a}^H_0(\theta_k)\text{Re}\left\{\bar{\mathbf{Z}}_0\right\}^{-1}\mathbf{a}_0(\theta_k),
\end{align}
where $\mathbf{\bar{Z}}_{\text{approx}} = \frac{1}{R_{\text{loss}} + R_i}\mathbf{Z}_{\text{approx}}$ and $\mathbf{\bar{Z}}_0 = \frac{1}{R_{\text{loss}} + R_i}\mathbf{Z}_0$. As seen from~\eqref{eq:signal_power_nula}, the power gain of the \ac{NULA} is $N_g$ times the gain of a superdirective antenna group. In the sequel, we determine $\mathbf{a}^H_0(\theta_k)\text{Re}\left\{\bar{\mathbf{Z}}_0\right\}^{-1}\mathbf{a}_0(\theta_k)$ in closed-form for $\bar{N}=2$. We first have that
\begin{equation}
\text{Re}\left\{\bar{\mathbf{Z}}_0\right\} =\begin{bmatrix}
1  & \bar{R}_m \\
\bar{R}_m & 1
\end{bmatrix},
\end{equation}
where $\bar{R}_m = \frac{1}{R_{\text{loss}} + R_i}R_m$. Then, the inverse matrix is determined as
\begin{equation}
\text{Re}\left\{\bar{\mathbf{Z}}_0\right\}^{-1} = \frac{1}{1 - \bar{R}^2_m}\begin{bmatrix}
1  & -\bar{R}_m \\
-\bar{R}_m & 1
\end{bmatrix},
\end{equation}
and
\begin{align}\label{eq:real_Z_inv_a}
\text{Re}\left\{\bar{\mathbf{Z}}_0\right\}^{-1}\mathbf{a}_0(\theta_k) &= \frac{1}{1 - \bar{R}^2_m}\begin{bmatrix}
1  & -\bar{R}_m \\
-\bar{R}_m & 1
\end{bmatrix}\!\begin{bmatrix}
1  \\
e^{-j\kappa \bar{d}\cos\theta_k}
\end{bmatrix} \nonumber \\
&=   \frac{1}{1 - \bar{R}^2_m}\begin{bmatrix}
1 -  \bar{R}_m e^{-j\kappa \bar{d}\cos\theta_k} \\
-\bar{R}_{m} + e^{-j\kappa \bar{d}\cos\theta_k}
\end{bmatrix},  
\end{align}
which gives 
\begin{align}\label{eq:signal_power_subarray}
&\mathbf{a}^H_0(\theta_k)\text{Re}\left\{\bar{\mathbf{Z}}_0\right\}^{-1}\mathbf{a}_0(\theta_k) \nonumber \\
&=  \frac{1}{1 - \bar{R}^2_m}\left(1 -  \bar{R}_m e^{-j\kappa \bar{d}\cos\theta_k} - \bar{R}_m e^{j\kappa \bar{d}\cos\theta_k} + 1 \right) \nonumber \\
& = \frac{2}{1 - \bar{R}^2_m}\left(1 - \bar{R}_m\cos(\kappa \bar{d}\cos\theta_k)\right).
\end{align}
According to~\eqref{eq:signal_power_subarray}, $\mathbf{a}^H_0(\theta_k)\text{Re}\left\{\bar{\mathbf{Z}}_0\right\}^{-1}\mathbf{a}_0(\theta_k) \approx 2$ for large~$\bar{d}$, i.e., $\bar{R}_m\approx 0$, which corresponds to the conventional power gain of a two-element array.
\begin{figure*}[t]
	\centering
	\begin{subfigure}{0.5\textwidth}
		\centering
		\includegraphics[width=1.12\linewidth]{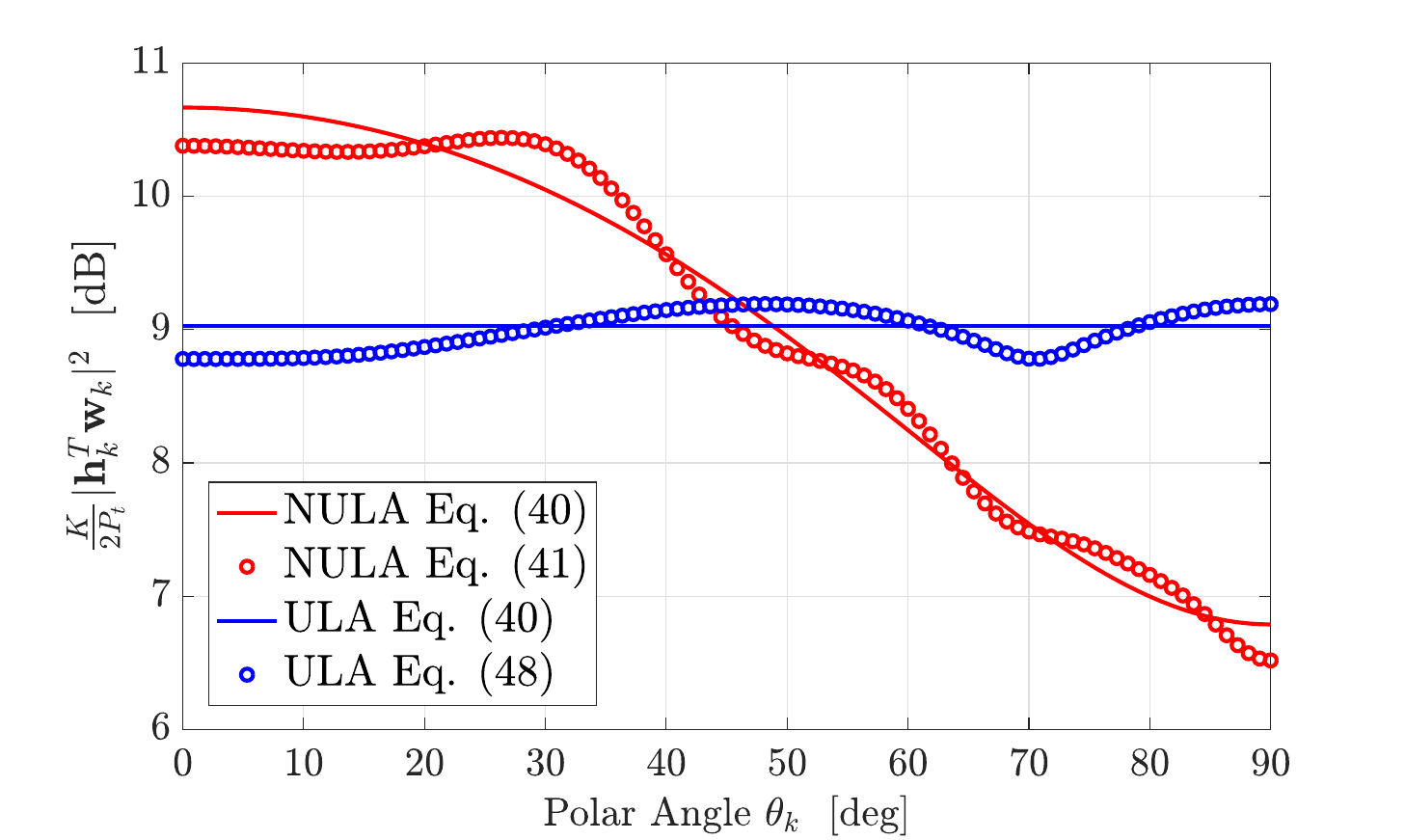}
		\caption{\footnotesize $d=1.5\lambda$, $d_g = 1.5\lambda$, and $\bar{d} = \lambda/5$.}
		\label{fig:Fig3a}
	\end{subfigure}~
	\begin{subfigure}{0.5\textwidth}
		\centering
		\includegraphics[width=1.1\linewidth]{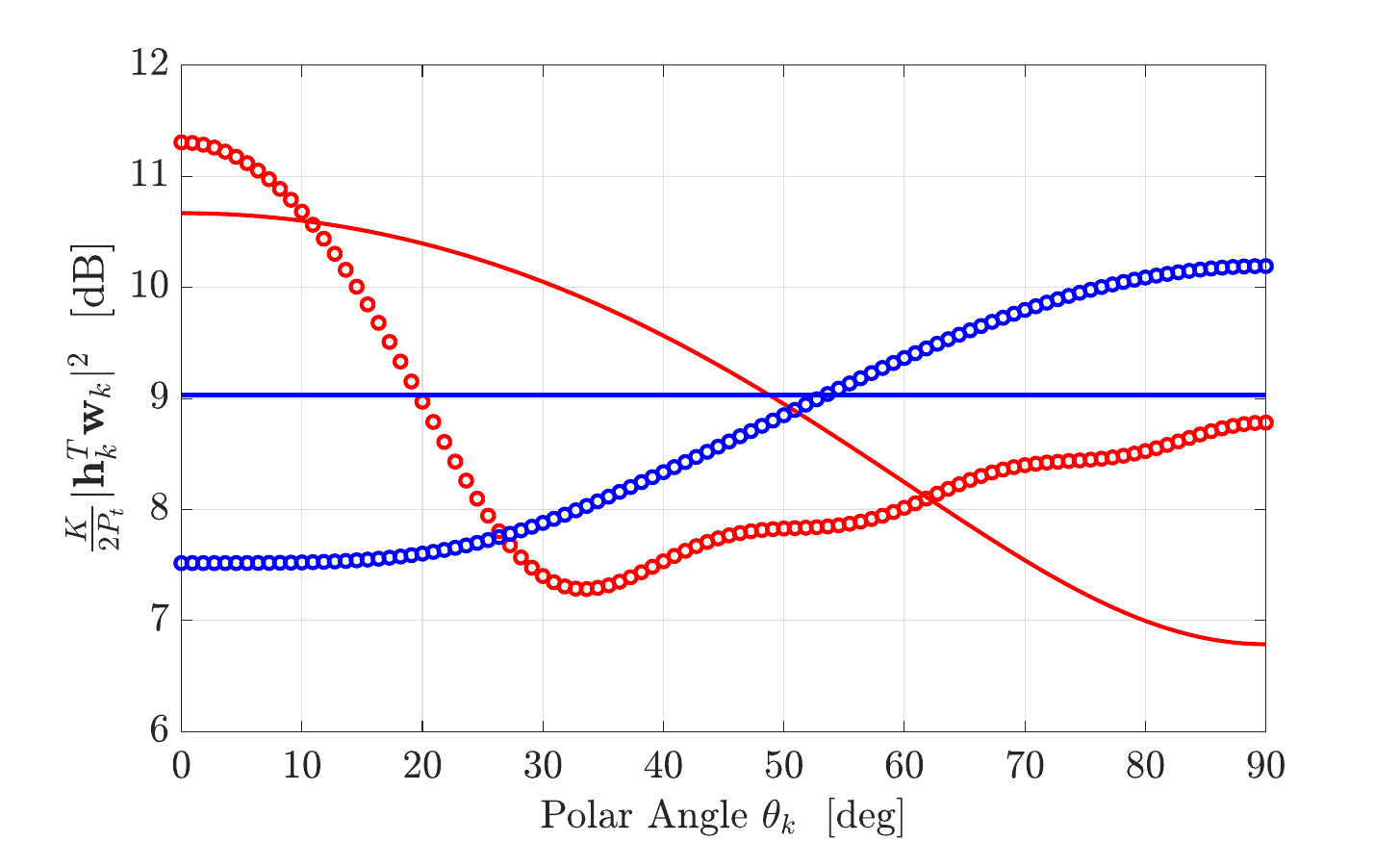}
		\caption{\footnotesize $d=\lambda/2$, $d_g = \lambda/2$, and $\bar{d} = \lambda/5$.}
		\label{fig:Fig3b}
	\end{subfigure}%
	\caption{Normalized signal power versus polar angle for $N = 8$ dipoles of $\ell=\lambda/2$, $\rho = \lambda/500$, and $\sigma_c=5.7\times 10^7$ S/m. In \ac{NULA}, $\bar{N}=2$ and $N_g = 4$. The carrier frequency is set to $f= 300$ GHz.}
	\label{fig:Fig3}
\end{figure*}

In a similar manner, the interference power at user~$k$ from the beam toward user $i\neq k$, normalized by $K/(2P_t)$, is given by 
\begin{align}\label{eq:interf_power}
\frac{K}{2P_t}&\left|\mathbf{h}_k^T\mathbf{w}_i \right|^2 = \frac{\left|\mathbf{a}^H(\theta_k)\text{Re}\left\{\bar{\mathbf{Z}}\right\}^{-1}\mathbf{a}(\theta_i)\right|^2}{\mathbf{a}^H(\theta_i)\text{Re}\left\{\bar{\mathbf{Z}}\right\}^{-1}\mathbf{a}(\theta_i)} \nonumber \\
& \approx \frac{\left|\mathbf{a}^H(\theta_k)\text{Re}\left\{\bar{\mathbf{Z}}_{\text{approx}}\right\}^{-1}\mathbf{a}(\theta_i)\right|^2}{\mathbf{a}^H(\theta_i)\text{Re}\left\{\bar{\mathbf{Z}}_{\text{approx}}\right\}^{-1}\mathbf{a}(\theta_i)} \nonumber \\[0.1cm]
&\overset{(a)}{=}  N_g\left|D_{N_g}\left(\kappa (d_g + (\bar{N}-1)\bar{d})(\cos\theta_k-\cos\theta_i)\right)\right|^2 \nonumber \\[0.1cm]
& \times \frac{\left|\mathbf{a}^H_0(\theta_k)\text{Re}\left\{\bar{\mathbf{Z}}_0\right\}^{-1}\mathbf{a}_0(\theta_i)\right|^2}{\mathbf{a}^H_0(\theta_i)\text{Re}\left\{\bar{\mathbf{Z}}_0\right\}^{-1}\mathbf{a}_0(\theta_i)},
\end{align}
where $(a)$ is proven in Appendix C. From~\eqref{eq:real_Z_inv_a} and~\eqref{eq:signal_power_subarray}, we finally have that 
\begin{align}
&\frac{\left|\mathbf{a}^H_0(\theta_k)\text{Re}\left\{\bar{\mathbf{Z}}_0\right\}^{-1}\mathbf{a}_0(\theta_i)\right|^2}{\mathbf{a}^H_0(\theta_i)\text{Re}\left\{\bar{\mathbf{Z}}_0\right\}^{-1}\mathbf{a}_0(\theta_i)} = \nonumber \\[0.1cm]
&=\frac{\left|1 + e^{j\kappa\bar{d}(\cos\theta_k-\cos\theta_i)} - \bar{R}_m\left(e^{j\kappa\bar{d}\cos\theta_k}+e^{-j\kappa\bar{d}\cos\theta_i}\right)\right|^2}{2(1-\bar{R}_m^2)\left(1 - \bar{R}_m\cos(\kappa \bar{d}\cos\theta_i)\right)}.
\end{align}

\subsection{Comparison With Uncoupled ULA}\label{sec:comparison_ula}
In the massive \ac{MIMO} literature, it is customary to consider a uniform inter-element spacing~\cite{mMIMO_book}. Furthermore, mutual coupling is avoided by employing a sufficiently large $d$ so that $\text{Re}\left\{\bar{\mathbf{Z}}\right\}\approx \mathbf{I}_N$. In this case, the normalized power of the desired signal becomes
\begin{equation}\label{eq:signal_power_nocoupling}
\frac{K}{2P_t}\left|\mathbf{h}_k^T\mathbf{w}_k\right |^2 \approx N,
\end{equation}
whereas the normalized interference is 
\begin{align}
\frac{K}{2P_t}\left |\mathbf{h}_k^T\mathbf{w}_i\right|^2 &  \approx N |D_{N}(\kappa d(\cos\theta_k-\cos\theta_i))|^2.
\end{align}
From~\eqref{eq:signal_power_nocoupling}, it is evident that the power gain is independent of the steering direction $\theta_k$ in an uncoupled \ac{ULA}. This assumption is safely made for a large antenna separation, as showcased in Fig.~\ref{fig:Fig3}. Specifically, half-wavelength spacing does create coupling between directional antennas, and hence the input impedance matrix cannot be ignored.\footnote{The assumption of uncoupled elements under half-wavelength spacing holds only for isotropic radiators~\cite{physical_modeling_com_systems}.} Lastly, according to the approximate expressions, the proposed \ac{NULA} with $\bar{N}=2$ yields a \textit{relative} power gain equal to 
\begin{align}
&N_g\mathbf{a}^H_0(\theta_k)\text{Re}\left\{\bar{\mathbf{Z}}_0\right\}^{-1}\mathbf{a}_0(\theta_k) - N = \nonumber \\ &=N_g\left[\mathbf{a}^H_0(\theta_k)\text{Re}\left\{\bar{\mathbf{Z}}_0\right\}^{-1}\mathbf{a}_0(\theta_k) - 2\right] = O(N_g),
\end{align}
which scales linearly with $N_g$ for some $\theta_k\in[0, \theta_{\max}]$; for example, $\theta_{\max}\approx 50\degree$ in Fig.~\ref{fig:Fig3}(\subref{fig:Fig3a}). This trade off between directivity and angular coverage is fundamental~\cite{antennas_and_prop_mimo}. Thus, increasing the former inevitably decreases the latter, and vice versa. Whether it is beneficial to have a highly directive array depends on the propagation environment and deployment scenario~\cite{antennas_and_prop_mimo,improve_mimo_directive_antennas}.
\begin{figure*}[t]
	\centering
	\begin{subfigure}{0.5\textwidth}
		\centering
		\includegraphics[width=0.92\linewidth]{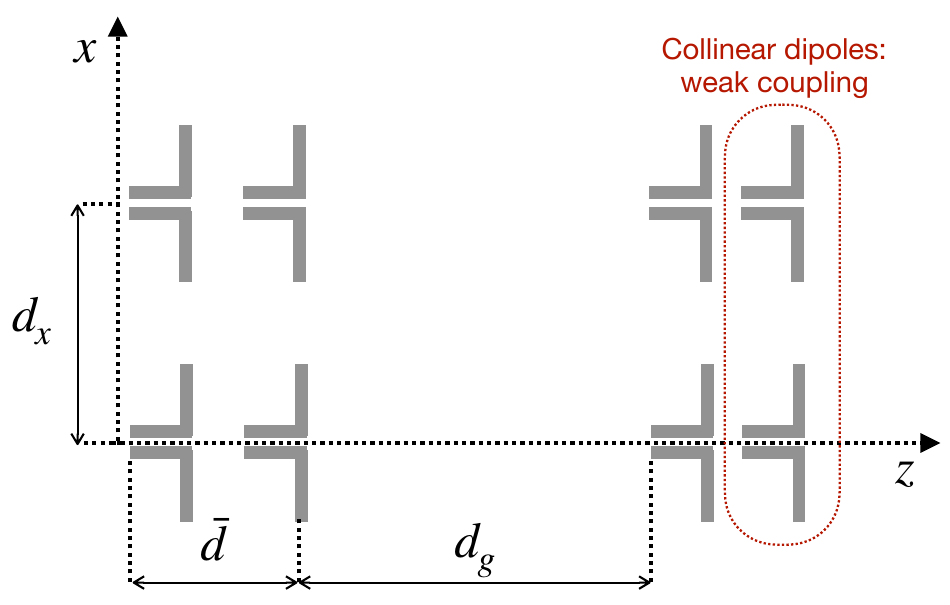}
		\caption{\footnotesize}
		\label{fig:Fig55a}
	\end{subfigure}~
	\begin{subfigure}{0.5\textwidth}
		\centering
		\includegraphics[width=0.72\linewidth]{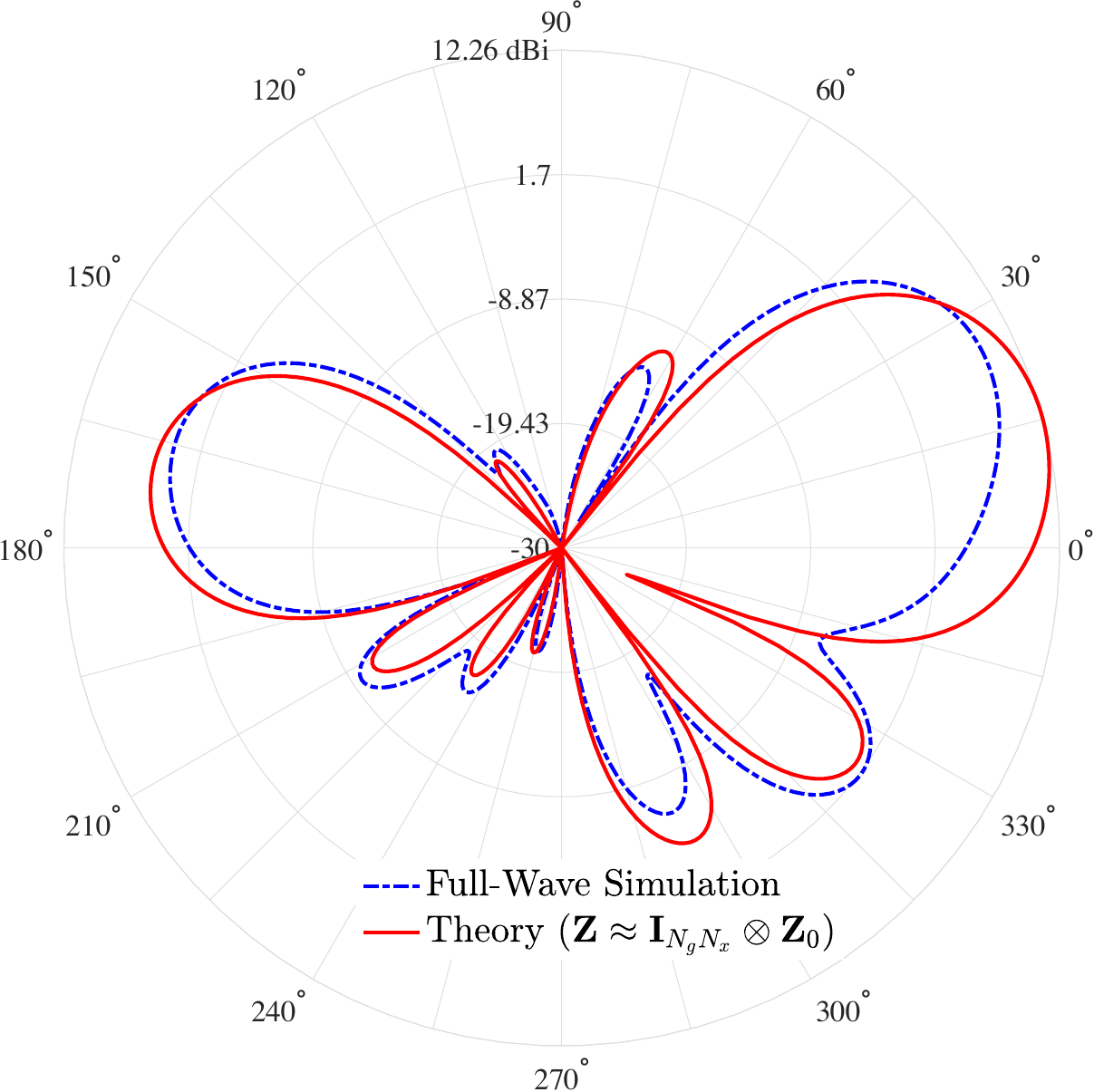}
		\caption{\footnotesize}
		\label{fig:Fig55b}
	\end{subfigure}
	\caption{(a) Example of a \ac{NUPA} with $N_x=N_g=\bar{N} = 2$; and (b) theoretical beampattern \eqref{eq:interf_power_nupa} against full-wave simulation for $d_x=0.7\lambda$, $d_g=1.5\lambda$, and $\bar{d}=\lambda/5$. The desired beamforming direction is $(\theta_i,\phi_i) = (20\degree,0\degree)$.}
	\label{fig:Fig55}
\end{figure*}

\subsection{Extension to Planar Arrays}\label{sec:planar_case}
In this section, we extend the proposed design to planar arrays, which are of practical importance. Specifically, a \ac{NUPA} is formed by placing $N_x$ \ac{NULA}s along the $x$-axis with inter-element spacing $d_x$, as depicted in Fig.~\ref{fig:Fig55}(\subref{fig:Fig55a}). The total number of antennas is $N=N_xN_g\bar{N}$. Also, the response vector of the \ac{NUPA} is given by $\mathbf{a}(\theta,\phi) = \mathbf{a}_x(\theta,\phi)\otimes  \mathbf{a}_g(\theta)\otimes \mathbf{a}_0(\theta)$, where $\mathbf{a}_x(\theta,\phi) \triangleq [1,\dots, e^{-j\kappa(N_x-1)d_x\cos\phi\sin\theta}]^T\in\mathbb{C}^{N_x\times 1}$. The coupling between the \ac{NULA}s is very small for $d_x\geq \lambda/2$ since adjacent dipoles along the $x$-axis are collinear. Given that, the impedance matrix of the \ac{NUPA} reduces to $\mathbf{Z}\approx \mathbf{I}_{N_gN_x}\otimes \mathbf{Z}_0$, while 
the signal and interference powers in \eqref{eq:signal_power_coupling} and \eqref{eq:interf_power} are recast, respectively, as
\begin{equation}\label{eq:signal_power_nupa}
\frac{K}{2P_t}\left|\mathbf{h}_k^T\mathbf{w}_k\right|^2 \approx N_x N_g \mathbf{a}^H_0(\theta_k)\text{Re}\left\{\bar{\mathbf{Z}}_0\right\}^{-1}\mathbf{a}_0(\theta_k),
\end{equation}
and
\begin{align}\label{eq:interf_power_nupa}
&\frac{K}{2P_t}\left|\mathbf{h}_k^T\mathbf{w}_i \right|^2 \approx \nonumber \\
&\approx N_xN_g
\left|D_{N_x}\left(\kappa d_x(\cos\phi_k\sin\theta_k-\cos\phi_i\sin\theta_i)\right)\right|^2\nonumber \\
& \times
\left|D_{N_g}\left(\kappa (d_g + (\bar{N}-1)\bar{d})(\cos\theta_k-\cos\theta_i)\right)\right|^2 \nonumber \\[0.1cm]
& \times \frac{\left|\mathbf{a}^H_0(\theta_k)\text{Re}\left\{\bar{\mathbf{Z}}_0\right\}^{-1}\mathbf{a}_0(\theta_i)\right|^2}{\mathbf{a}^H_0(\theta_i)\text{Re}\left\{\bar{\mathbf{Z}}_0\right\}^{-1}\mathbf{a}_0(\theta_i)}.
\end{align}
From~\eqref{eq:signal_power_nupa}, we see that the signal power is $N_x$ times that of a \ac{NULA} with $N_g$ dipole pairs, as expected. Regarding the interference power, \ac{NUPA} offers an additional degree of freedom compared to a single \ac{NULA} along the $z$-axis, which is given by the term $N_x
\left|D_{N_x}\left(\kappa d_x(\cos\phi_k\sin\theta_k-\cos\phi_i\sin\theta_i)\right)\right|^2$. The good accuracy of \eqref{eq:interf_power_nupa} is confirmed in Fig.~\ref{fig:Fig55}(\subref{fig:Fig55b}) considering the polar plane $\phi_k=\phi_i = 0$. Note that the full-wave simulation was performed using the Antenna Toolbox of MATLAB.

\section{Channel Estimation}\label{sec:channel_est}
So far, we have assumed perfect channel knowledge at the \ac{BS} and analyzed the performance of \ac{MRT} in the presence of mutual coupling. In this section, we focus on the channel estimation problem, which is crucial to beamforming. 

\subsection{Channel Reciprocity}
In massive \ac{MIMO}, it is typical to invoke channel reciprocity for \ac{TDD} operation~\cite{massive_mimo_tdd}. This enables the \ac{BS} to estimate the downlink channel through uplink pilots sent by users. Let $\mathbf{h}_{k,\text{DL}} = \mathbf{h}_k^T\in \mathbb{C}^{1\times N}$ denote the downlink channel, where $ \mathbf{h}_k$ is given by~\eqref{eq:channel_vector}. The \ac{TDD} assumption is then that $\mathbf{h}_{k,\text{UL}} = \mathbf{h}^T_{k,\text{DL}}$. Although the physical channels (i.e., those defined by electromagnetic theory) are reciprocal, this does not generally hold for their information-theoretic counterparts in the presence of antenna mutual coupling~\cite{tdd_reciprocity_coupling}. In particular, the \ac{BS} needs to employ a linear transformation to compute $ \mathbf{h}^T_{k,\text{DL}}$ from the estimated $ \mathbf{h}_{k,\text{UL}}$. Nevertheless, the transmit and receive array gains coincide with each other under isotropic background noise and noise matching at the receiver~\cite[Eq. (97)]{circuit_th_commun}. Thus, we can assume that $\mathbf{h}_{k,\text{UL}} = \mathbf{h}^T_{k,\text{DL}}$, and that the reception strategy maximizing the \ac{SNR} of each user $k$ is the maximum ratio combiner $\mathbf{v}_k = \mathbf{h}_k^H/\|\mathbf{h}_k\|$.

\subsection{Problem Formulation}
We assume a block-fading model, where the channel coherence time is much larger than the training period. The \ac{BS} estimates the uplink channel $\mathbf{h}_k$ of each user $k$ in rounds. Subsequently, we focus on an arbitrary user and omit the subscript ``$k$''. Specifically, the training period for each user consists of $N_{\text{slot}}$ time slots. At each time slot $t= 1,\dots, N_{\text{slot}}$, the user transmits the pilot signal $x_t = \sqrt{P_p}$, where $P_p$ is the power per pilot signal. In turn, the \ac{BS} combines the received pilot signal using a training hybrid combiner $\mathbf{V}_t=\mathbf{V}_{\text{RF},t}\mathbf{V}_{\text{BB},t} \in\mathbb{C}^{N\times N_{\text{RF}}}$. Therefore, the post-processed signal at slot $t$, $\mathbf{y}_t\in\mathbb{C}^{K\times 1}$, is written as
\begin{equation}
\mathbf{y}_t = \sqrt{\beta P_p} \mathbf{V}_t^H\mathbf{h} + \mathbf{V}_t^H\mathbf{n}_t,
\end{equation}
where $\mathbf{n}_t\sim\mathcal{CN}(\mathbf{0}_{N\times 1}, \sigma^2\mathbf{I}_{N})$ is the additive noise vector. Let $N_{\text{beam}} = N_{\text{slot}}N_{\text{RF}}$ denote the total number of pilot beams. After $N_{\text{slot}}$ training slots, the BS acquires the measurement vector $\bar{\mathbf{y}} \triangleq [\mathbf{y}^T_1, \dots, \mathbf{y}^T_{N_{\text{slot}}}]^T \in\mathbb{C}^{N_{\text{beam}} \times 1}$ for $\mathbf{h}$ as
\begin{align}\label{eq:general_expression_training}
\bar{\mathbf{y}}  &= 
\sqrt{\beta P_p}\begin{bmatrix}
\mathbf{V}^H_1\\
\vdots \\
\mathbf{V}^H_{N_{\text{slot}}}
\end{bmatrix} \mathbf{h} 
+
\begin{bmatrix}
\mathbf{V}^H_1\mathbf{n}_1\\
\vdots \\
\mathbf{V}^H_{N_{\text{slot}}}\mathbf{n}_{N_{\text{slot}}}
\end{bmatrix}  
= \sqrt{\beta P_p}\ \overline{\mathbf{V}}^H \mathbf{h} + \bar{\mathbf{n}},
\end{align}
where $\overline{\mathbf{V}} \triangleq [\mathbf{V}_1, \dots, \mathbf{V}_{N_\text{slot}}]\in\mathbb{C}^{N\times N_{\text{beam}}}$, whereas $\bar{\mathbf{n}}\in\mathbb{C}^{N_{\text{beam}}\times 1}$ is the effective noise. More particularly, $\mathbf{R}_{\bar{\mathbf{n}}[s]} \triangleq \sigma^2\text{blkdiag}\left(\mathbf{V}^H_1\mathbf{V}_1,\dots, \mathbf{V}^H_{N_\text{slot}}\mathbf{V}_{N_\text{slot}}\right)$ is the covariance matrix of the effective noise, which is colored in general.\footnote{We have assumed the same noise variance, $\sigma^2$, as in the user side.} Regarding the pilot combiners, due to the hybrid array architecture, $\overline{\mathbf{V}} = \overline{\mathbf{V}}_{\text{RF}}\overline{\mathbf{V}}_{\text{BB}}$, with $\overline{\mathbf{V}}_{\text{RF}}=[\mathbf{V}_{\text{RF},1}, \dots, \mathbf{V}_{\text{RF}, N_{\text{slot}}}]\in\mathbb{C}^{N\times N_{\text{beam}}}$ and $\overline{\mathbf{V}}_{\text{BB}} = \text{blkdiag}(\mathbf{V}_{\text{BB},1}, \dots, \mathbf{V}_{\text{BB}, N_{\text{slot}}})\in\mathbb{C}^{N_{\text{beam}}\times N_{\text{beam}}}$ comprising the pilot \ac{RF} beams and baseband combiners of the $N_{\text{slot}}$ time slots, respectively.

\subsection{Orthogonal Matching Pursuit}\label{sec:omp}
\subsubsection{Sparse Formulation}
We next consider a dictionary $\bar{\mathbf{H}}\in~\mathbb{C}^{N\times G}$ whose $G$ columns are the channel vectors associated with a predefined set of \ac{AoA}. Then, the uplink channel can be approximated as
\begin{equation}
\mathbf{h} \approx  \bar{\mathbf{H}}\bm{\beta},
\end{equation}
where $\bm{\beta}$ is a $G\times 1$ vector with a single nonzero entry corresponding to the \ac{LoS} path. Therefore,~\eqref{eq:general_expression_training} is recast as $\bar{\mathbf{y}} \approx \mathbf{\Phi} \bm{\beta}  + \bar{\mathbf{n}}$, where $\mathbf{\Phi}\triangleq \sqrt{P_p}\ \overline{\mathbf{V}}^H\bar{\mathbf{H}} \in\mathbb{C}^{N_{\text{beam}}\times G}$ is the \textit{equivalent} sensing matrix. Since $G\gg 1$, the vector $\bar{\bm{\beta}}$ is $1$-sparse, and thus the channel estimation problem can be formulated as the sparse recovery problem~\cite{jsac_2021}
\begin{align}\label{eq:opt_problem}
\hat{\bm{\beta}} = \arg&\min_{\bm{\beta}} \  \|\bm{\beta} \|_1 \quad \text{subject to} \quad \left \| \bar{\mathbf{y}} -\mathbf{\Phi} \bm{\beta}\right \| \leq \epsilon,
\end{align}
where $\epsilon \leq \mathbb{E}\{\|\bar{\mathbf{n}}\|\}$ is an appropriately chosen bound on the mean magnitude of the effective noise. The $l_1$-norm optimization problem in~\eqref{eq:opt_problem} can be readily solved by the popular \ac{OMP} algorithm, which takes the following form for single-path channels
\begin{align}
g^\star = \underset{g\in\mathcal{G}}{\arg\max} \  \left|\mathbf{\Phi}^H(g)\bar{\mathbf{y}}\right |,
\end{align}
where $\mathcal{G}$ denotes the set of predefined \ac{AoA}. Finally, the estimate of $\mathbf{h}$ is obtained as $\hat{\mathbf{h}}=\bar{\mathbf{H}}(g^{\star})$. It is worth stressing that the actual \ac{AoA} might differ from the one defined by the dictionary. Nonetheless, this mismatch error can become negligible by adopting a high-resolution dictionary, as demonstrated in~\cite{jsac_2021}.
\begin{figure}[t]
	\centering
	\includegraphics[width=0.8\linewidth]{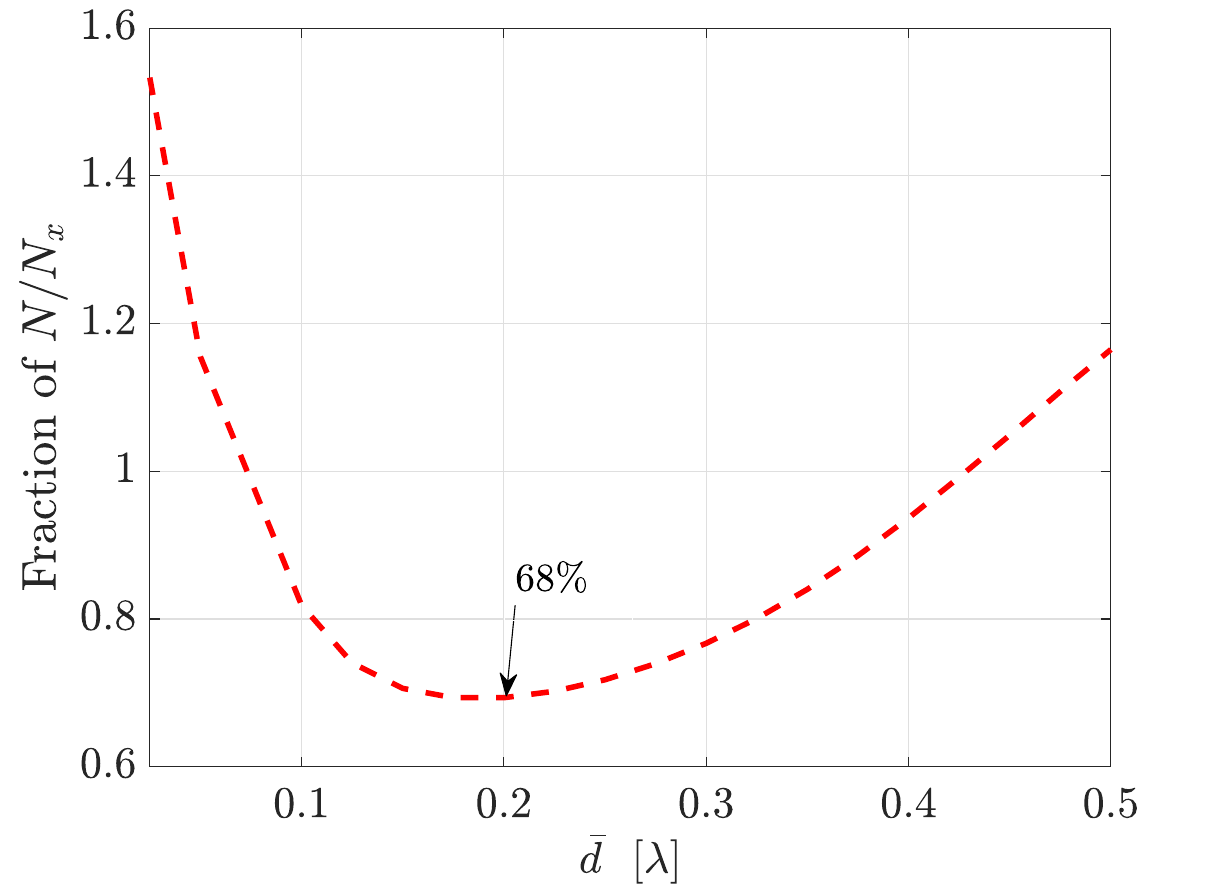}
	\caption{Fraction of $N/N_x$ versus $\bar{d}$ for $(\theta,\phi)=(0,0)$ at $f=300$~GHz. The dipoles are half-wavelength copper wires of radius $\rho=\lambda/500$.}
	\label{fig:Fig44}
\end{figure}
\subsubsection{Dictionary and Pilot Beams}
In the spirit of~\cite{mmwave_omp1,mmwave_omp3}, we discretize the polar angle $\theta\in [0,\theta_{\max}]$ and azimuth angle $\phi\in [0,\phi_{\max}]$ as 
\begin{align}
\bar{\theta}_{g_z} &= \frac{\theta_{\max}}{G_z}g_z, \quad g_z=0,\dots, G_z-1,\\
\bar{\phi}_{g_x} &= \frac{\phi_{\max}}{G_x}g_x, \quad g_x=0,\dots, G_x-1,
\end{align}
where $G=G_xG_z$ is the overall dictionary size, which results in the coupling-aware dictionary
\begin{equation}
\bar{\mathbf{H}} = \left[\text{Re}\{\mathbf{Z}\}^{-\frac{1}{2}}\mathbf{a}\left(\bar{\theta}_0,\bar{\phi}_0\right), \dots, \text{Re}\{\mathbf{Z}\}^{-\frac{1}{2}}\mathbf{a}\left(\bar{\theta}_{G_z-1},\bar{\phi}_{G_x-1}\right) \right].
\end{equation}
The elements of the \ac{RF} combiner $\overline{\mathbf{V}}_{\text{RF}}$ are selected from the set $\{-1/\sqrt{N},1/\sqrt{N}\}$ with equal probability. The reason we adopt a randomly formed RF combiner is that it will exhibit low mutual-column coherence, and therefore is expected to attain a high recovery probability according to the \ac{CS} theory~\cite{omp_algorithm}. The columns of $\overline{\mathbf{V}}_{\text{RF}}$ have been normalized so that the total power consumed during the channel estimation stage is $P_{\text{CE}} = K N_\text{beam}P_p$, as $KN_\text{beam}$ pilot beams are used for the $K$ users. The specific RF pilot design results in a colored effective noise. For this reason, we design the baseband combiner such that the effective noise remains white. Let $\mathbf{D}^H_t\mathbf{D}_t$ be the Cholesky decomposition of $\mathbf{V}^H_{\text{RF},t}\mathbf{V}_{\text{RF},t}$, where $\mathbf{D}\in\mathbb{C}^{N_{\text{RF}}\times N_{\text{RF}}}$ is an upper triangular matrix. Then, the baseband combiner of the $t$th slot is set to $\mathbf{V}_{\text{BB},t} = \mathbf{D}^{-1}_t$, and hence $\overline{\mathbf{V}} = \overline{\mathbf{V}}_{\text{RF}}\text{blkdiag}(\mathbf{D}^{-1}_1,\dots, \mathbf{D}^{-1}_{N_{\text{slot}}})$. Under this pilot beam design, the covariance matrix of the effective noise becomes $\mathbf{R}_{\bar{\mathbf{n}}} = \sigma^2\mathbf{I}_{N_{\text{beam}}}$.
 \begin{figure*}[t]
 	\centering
 	\begin{subfigure}{0.5\textwidth}
 		\centering
 		\includegraphics[width=0.93\linewidth]{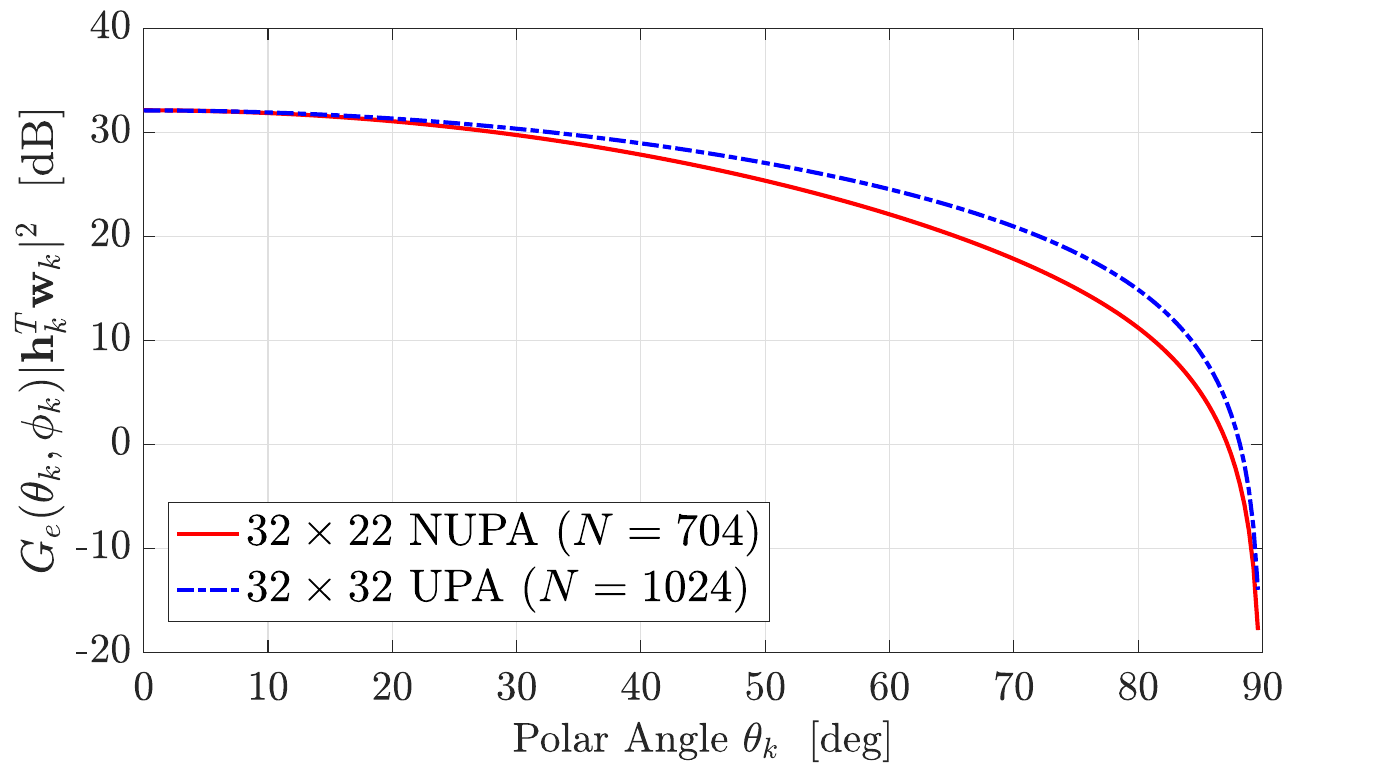}
 	\end{subfigure}~
 	\begin{subfigure}{0.5\textwidth}
 		\centering
 		\includegraphics[width=0.93\linewidth]{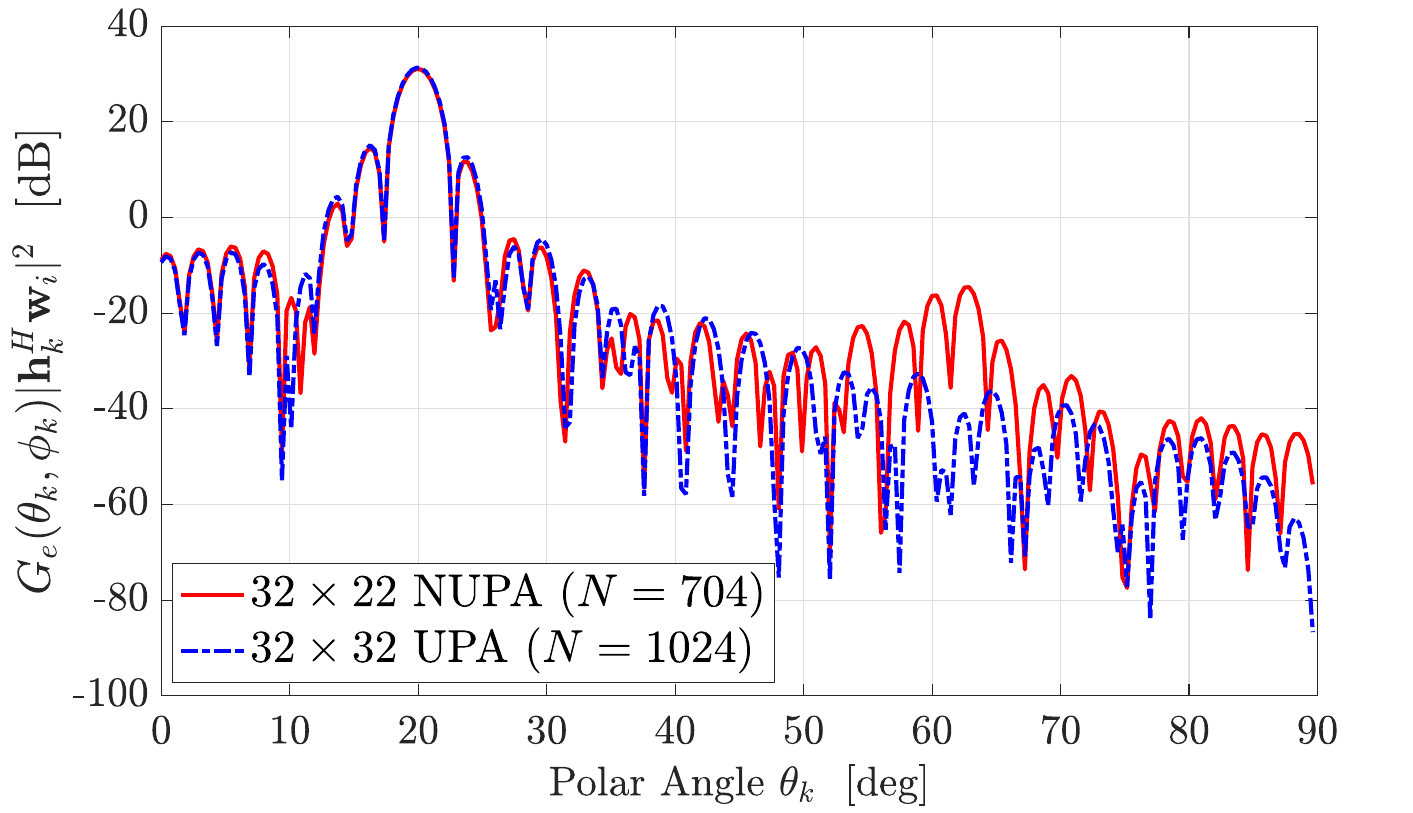}
 	\end{subfigure}
 	\caption{Signal and interference powers. In \ac{NUPA}, $\bar{N}=2$, $N_g = 11$, $d_x=0.7\lambda$, $\bar{d} = \lambda/5$, and $d_g=1.95\lambda$. In \ac{UPA}, $d=d_x=0.7\lambda$. The interference power is computed for $(\theta_i,\phi_i) = (20\degree,0\degree)$. Also, $P_t=1/2$, $K=1$, and $f= 300$ GHz.}
 	\label{fig:Fig4}
 \end{figure*}
\begin{figure*}[t]
	\centering
	\includegraphics[width=1.02\linewidth]{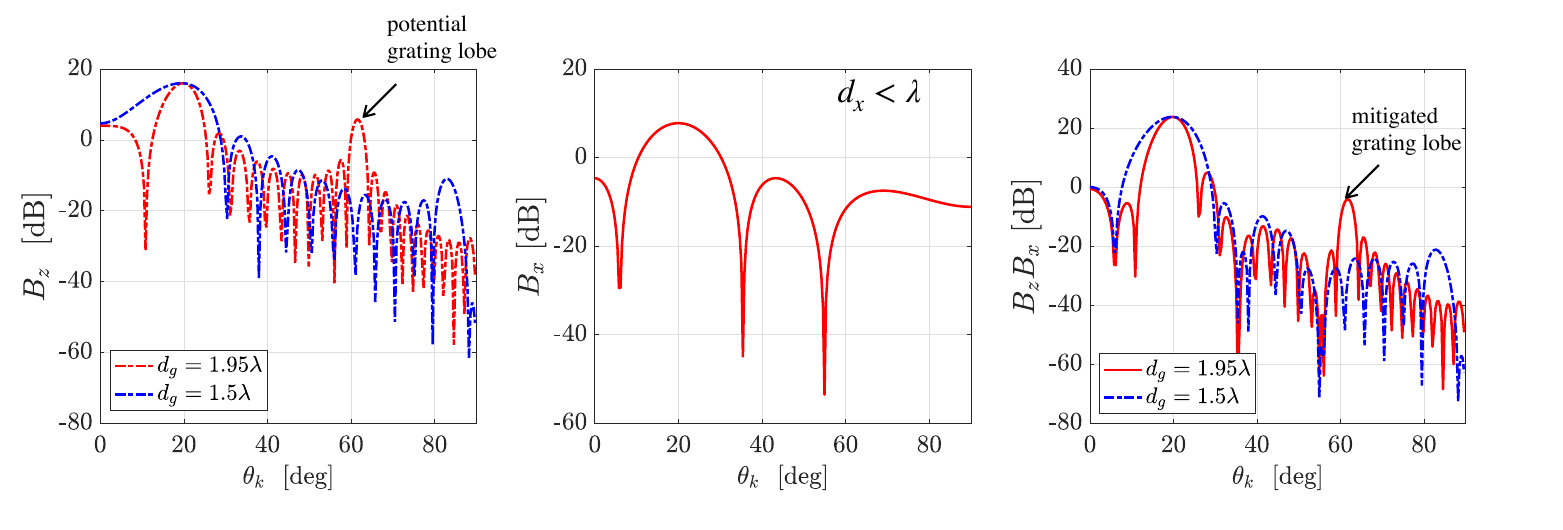}
	\caption{Polar beampatterns of a $6\times 22$-element \ac{NUPA} with $\bar{d}=\lambda/5$, $d_x=0.7\lambda$, and various $d_g$. The beamforming angle is~$(\theta_i,\phi_i)= (20\degree,0\degree)$.}
	\label{fig:Fig45}
\end{figure*}
\section{Reducing the Number of BS Antennas}\label{sec:reduced_num_antennas}
As demonstrated in Section~\ref{sec:comparison_ula}, the proposed \ac{NULA} achieves larger gain than \ac{ULA} thanks to the superdirective pairs. This excessive power gain can improve \ac{EE}, because the number of antennas is kept fixed in both array designs. A more radical approach is to adopt a \ac{NULA} with fewer elements than a \ac{ULA} to substantially reduce the power consumption, akin to the paradigm of array thinning. Hereafter, we focus on the general case of planar arrays, and determine the parameters $N_g$, $d_g$, and $\bar{d}$ of \ac{NUPA} in order to attain a similar signal and interference power as a \ac{UPA} of $N$ elements. 

\subsection{How Many BS Antennas Do We Need?}\label{sec:how_many_antennas}
For the sake of fair comparison, we consider that both \ac{NUPA} and \ac{UPA} have $N_x$ elements along the $x$-axis with inter-element spacing $d_x$. We then seek to find how many antennas along the $z$-axis are needed such that the two arrays offer the same signal power toward the endfire direction $(\theta,\phi)=(0\degree,0\degree)$. This occurs when
\begin{equation}
N_xN_g\mathbf{a}_0^H(0)\text{Re}\left\{\bar{\mathbf{Z}}_0\right\}^{-1}\mathbf{a}_0(0) = N, 
\end{equation}
which gives, after some basic algebra,
\begin{equation}\label{eq:opt_num_antennas}
2N_g = \frac{1 - \bar{R}^2_m}{\left(1-\bar{R}_m\cos\left(\kappa\bar{d}\right)\right)}\frac{N}{N_x}.
\end{equation}
Note that \eqref{eq:opt_num_antennas} hinges on the inter-element spacing $\bar{d}$ within dipole pairs, which determines the level of mutual coupling. From Fig.~\ref{fig:Fig44}, we observe that the minimum number of antennas is attained for $\bar{d}=\lambda/5$.\footnote{For spacings smaller than $\lambda/5$, the ohmic losses become dominant and decrease the array gain.} In this case, the \ac{NUPA} requires approximately $68\%$ of the \ac{UPA} antennas along the $z$-axis, and hence up to $32\%$ saving in \ac{RF} hardware is possible. This performance can be readily obtained for a point-to-point link where the user is placed at the endfire direction of the \ac{BS}. Conversely, the impact of reducing the number of \ac{BS} antennas on the performance of multiuser transmissions is hard to analytically study, under either perfect or imperfect \ac{CSI}. For this reason, we resort to numerical simulations in Section~\ref{sec:numerical_results}.

\subsection{Spatial Resolution and Inter-Group Spacing }
It is known that spatial resolution is determined by the array size~\cite{balanis_book}. Thus, the \ac{NUPA} with less elements will preserve its spatial resolution if its length along the $z$-axis is equal to that of \ac{UPA}. This happens for $N_g\bar{d} + (N_g-1)d_g = (N/N_x-1)d$, or equivalently
\begin{equation}
d_g = \frac{(N/N_x-1)d - N_g\bar{d}}{N_g-1},
\end{equation}
which ensures that the \ac{NUPA} and \ac{UPA} have the same physical size. To demonstrate this design methodology, we consider a $32\times 32$-element \ac{UPA} at the \ac{BS} with $d=d_x=0.7\lambda$. Then, the \ac{NUPA} will consist of $N_g = 11$ dipole pairs, yielding a $32\times 22$-element array with inter-group spacing $d_g = 1.95\lambda$. We now compare these two arrays in terms of the overall signal and interference powers given by $\beta_k\left| \mathbf{h}_k^T\mathbf{w}_k\right|^2 \propto  G_e(\theta_k,\phi_k)\left|\mathbf{h}_k^T\mathbf{w}_k\right|^2$ and $\beta_k \left|\mathbf{h}_k^T\mathbf{w}_i\right|^2 \propto  G_e(\theta_k,\phi_k)\left|\mathbf{h}_k^T\mathbf{w}_i\right|^2$, respectively. From Fig.~\ref{fig:Fig4}, we verify that both arrays have very similar performance, though the \ac{NUPA} employs much fewer elements than \ac{UPA}. 
\begin{figure*}[t]
	\centering
	\includegraphics[width=0.9\linewidth]{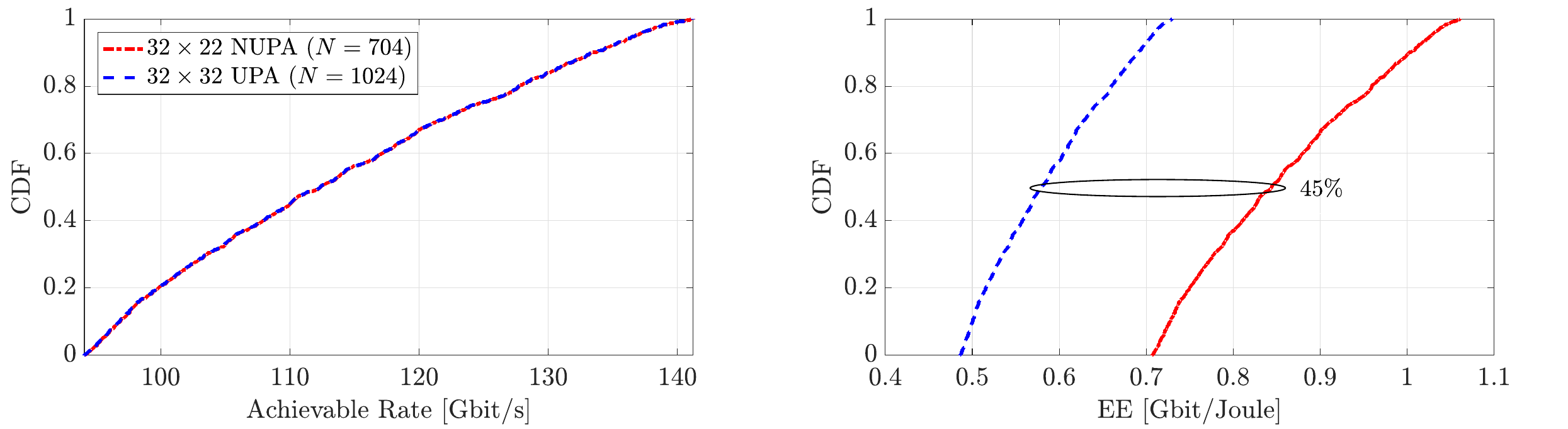}
	\caption{Results for a point-to-point link. In the \ac{NUPA}, $\bar{N}=2$, $N_g = 11$, and $\bar{d}=\lambda/5$.}
	\label{fig:Fig6}
\end{figure*}
Lastly, the side lobes occurred for polar angles $\theta_k$ beyond $50\degree$ can be neglected since their power is less than $0$ dB. About possible grating lobes, these will appear if $d_g+\bar{d}>\lambda$ and $d_x>\lambda$. Their positions, in the polar plane (i.e., $\phi_k=\phi_i=0$), are given by~\cite{balanis_book}
\begin{align}\label{eq:theta_k}
\theta_k = \cos^{-1}\left(\cos\theta_i \pm n_1 \frac{\lambda}{d_g + \bar{d}}\right) = \sin^{-1}\left(\sin\theta_i \pm n_2 \frac{\lambda}{d_x}\right), 
\end{align} 
for $n_1$ and $n_2$ in $\{1,2,3,\dots\}$. Since $d_x < \lambda$, \eqref{eq:theta_k} does not admit a real solution, and hence grating lobes do not occur in the visible region of the \ac{NUPA}. In practice, these are suppressed by the low-level sidelobes of the array along the $x$-axis, as indicated by \eqref{eq:interf_power_nupa}. This behavior is confirmed in Fig.~\ref{fig:Fig45}, where
\begin{align}
B_z &= G_e(\theta_k,\phi_k) \nonumber \\
&\times N_g\left|D_{N_g}\left(\kappa (d_g + (\bar{N}-1)\bar{d})(\cos\theta_k-\cos\theta_i)\right)\right|^2 \nonumber \\
& \times \frac{\left|\mathbf{a}^H_0(\theta_k)\text{Re}\left\{\bar{\mathbf{Z}}_0\right\}^{-1}\mathbf{a}_0(\theta_i)\right|^2}{\mathbf{a}^H_0(\theta_i)\text{Re}\left\{\bar{\mathbf{Z}}_0\right\}^{-1}\mathbf{a}_0(\theta_i)}
\end{align}
and
\begin{align}
B_x &= N_x 
\left|D_{N_x}\left(\kappa d_x(\cos\phi_k\sin\theta_k-\cos\phi_i\sin\theta_i)\right)\right|^2
\end{align}
are the beampatterns of the \ac{NUPA} along the $z$ and $x$ directions, respectively~\cite{grating_lobe_supp}.
\begin{remark}
The proposed \ac{NUPA} enables beam broadening without sacrificing the maximum array gain by adopting a small inter-group spacing $d_g$, i.e., by decreasing the length of the overall array along the $z$-axis. This feature is showcased in Fig.~\ref{fig:Fig45}, and can be very useful for \ac{THz} links where wide beamwidths alleviate the detrimental effect of beam misalignment~\cite{thz_misalignment}. 
\end{remark}

\begin{table}[H]
	\centering
	\caption{Main Simulation Parameters~\cite{dynamic_aosa_thz,thz_channel_model,thz_misalignment}}
	\label{Table:SimParameters}
	\small
	\begin{tabular}{|l  l|}
		\hline
		\textbf{Parameter} & \textbf{Value} \\
		\hline
		Carrier frequency, bandwidth & $f = 300$ GHz, $B=15$ GHz \\
		BS's input power & $P_t=20$ dBm \\
		Power per pilot symbol & $P_p = 20$ dBm \\
		Power density of noise& $\sigma^2 = -174$ dBm/Hz\\
		User distance  & $r_k\sim\mathcal{U}\left(5, 15\right)$ m \\
		Absorption coefficient & $k_{\text{abs}}=0.0033 \ \text{m}^{-1}$\\
		\hline 
		Dipole length, radius & $\ell = \lambda/2$, $\rho = \lambda/500$ \\
		Copper conductivity & $\sigma_c=5.7\times 10^7$ S/m\\
		\hline
		Baseband unit  & $P_{\text{BB}} = 200$ mW \\
		Phase shifter  & $P_{\text{PS}} = 42$ mW  \\
		Power amplifier & $P_{\text{PA}} =60$ mW \\
		DAC & $P_{\text{DAC}} =110$ mW \\
		Local oscillator & $P_{\text{LO}} =4$ mW\\
		Mixer &   $P_{\text{M}}=22$ mW\\
		\hline
	\end{tabular}
\end{table}

\section{Simulation Results}\label{sec:numerical_results}
We conduct extensive numerical simulations to assess the performance of the proposed array design. In all numerical experiments, we consider a fractional bandwidth $B/f \leq 0.1$ and \ac{LoS} links, which ensure a spatially narrowband propagation channel~\cite{jsac_2021}. The other simulation parameters are summarized in Table~\ref{Table:SimParameters}. Also, all values in the power consumption model are taken from~\cite{dynamic_aosa_thz}.

\subsection{Point-to-Point Link}
We commence with the case of a point-to-point link, where a single user is placed at $(\theta,\phi)=(0\degree,0\degree)$. Note that this setup can represent a wireless backhaul link. Due to the fixed user direction, perfect \ac{CSI} is assumed at the \ac{BS}. The distance from the \ac{BS} follows the uniform distribution $\mathcal{U}(5,15)$~m. The maximum achievable rate and \ac{EE} are given by $R = B\log_2\left(1 + \frac{\beta| \mathbf{h}^T\mathbf{w}|^2}{B\sigma^2}\right)$ and $R/P_c$, respectively. The signal power $|\mathbf{h}^T\mathbf{w}|^2$ is determined by \eqref{eq:signal_power_nupa} and is independent of $d_g$ and $d_x$. The \ac{CDF} of each metric is calculated for $1,000$ channel realizations. From Fig.~\ref{fig:Fig6} of the next page, we observe that the \ac{NUPA} attains the same achievable rate as \ac{UPA}, yet with $320$ antennas less. This translates roughly to $31\%$ saving in \ac{RF} circuitry, such as power amplifiers, phase shifters and combiners, which in turn yields a mean \ac{EE} improvement of $45\%$. 

\begin{figure*}[t]
	\centering
	\includegraphics[width=1.02\linewidth]{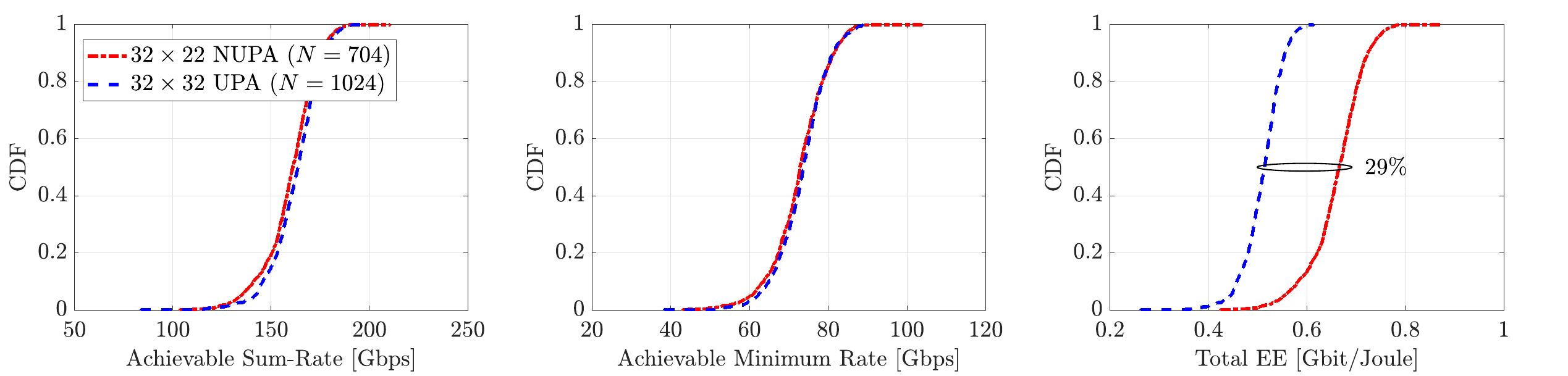}
	\caption{Results for $K=2$ users with directions $\theta_k\sim\mathcal{U}(0\degree,50\degree)$ and $\phi_k\sim\mathcal{U}(0\degree,360\degree)$. The \ac{BS} acquires \ac{CSI} with partial training of $N_{\text{beam}} = 0.8 N$ pilots per user. The dictionary size is $G=N$ for each array design. In \ac{NUPA}, $\bar{d}=\lambda/5$, $d_g=1.95\lambda$, and $d_x=0.7\lambda$. In \ac{UPA}, $d_x=d=0.7\lambda$.}
	\label{fig:Fig7}
\end{figure*}
\begin{figure}[t]
	\centering
	\includegraphics[width=0.65\linewidth]{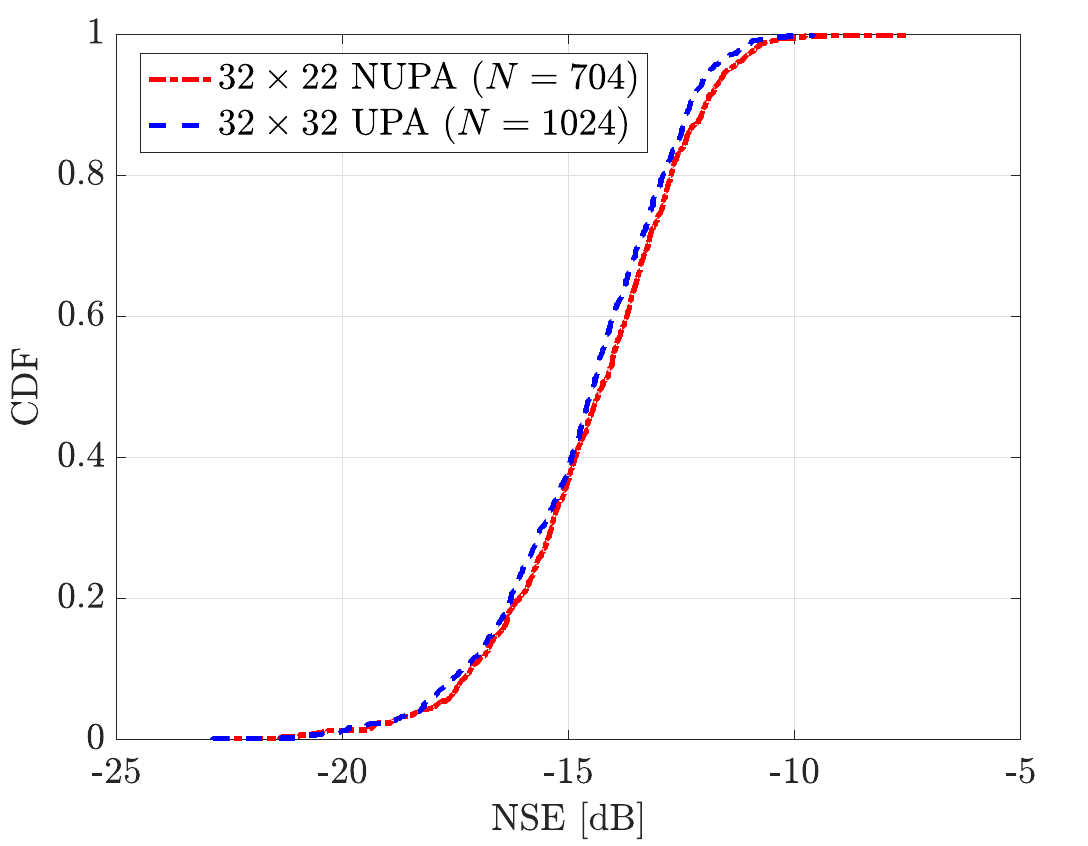}
	\caption{\ac{CDF} of the NSE for $K=2$ users and partial beam training with $N_{\text{beam}} = 0.8 N$ pilots per user.}
	\label{fig:Fig8}
\end{figure}	
For the point-to-point deployment, we can employ single-port impedance matching as the beamforming angle is kept fixed all the time. Subsequently, we investigate how this is accomplished. To calculate the active impedance of each dipole, the endfire current vector is decomposed as $\mathbf{i}^{\text{opt}} = [\mathbf{i}_0^{\text{opt}},\dots, \mathbf{i}^{\text{opt}}_{N_g-1}]^T$, where  $\mathbf{i}^{\text{opt}}_{n_g} \triangleq [I_{n_g}(0), I_{n_g+1}(0)]^T$ denotes the optimal excitation vector at each antenna pair. This is defined~as
\begin{align}\label{eq:current_decomposition}
\mathbf{i}^{\text{opt}}_{n_g} &= \frac{\sqrt{2P_t} e^{-j\kappa n_g(d_g + \bar{d})}}{\sqrt{N_g\mathbf{a}^H_0(0)\text{Re}\{\mathbf{Z}_0\}^{-1}\mathbf{a}_0(0)}}\text{Re}\{\mathbf{Z}_0\}^{-1}\mathbf{a}_0(0) \nonumber \\
&=\frac{\sqrt{P_t}e^{-j\kappa n_g(d_g + \bar{d})}}{\sqrt{N_g(R_{\text{self}}-R_m\cos(\kappa\bar{d}))}} \begin{bmatrix}
R_{\text{self}} -  R_m e^{-j\kappa \bar{d}} \\[0.2cm]
-R_{m} + R_{\text{self}} e^{-j\kappa \bar{d}}
\end{bmatrix},  
\end{align}
where \eqref{eq:current_decomposition} follows from~\eqref{eq:opt_current} and \eqref{eq:signal_power_subarray} for $\theta=0\degree$. Therefore, for any pair $n_g$, we have that 
\begin{equation}
\frac{I_{n_g+1}(0)}{I_{n_g}(0)} = \frac{-R_{m} + R_{\text{self}}e^{-j\kappa \bar{d}}}{R_{\text{self}} -  R_m e^{-j\kappa \bar{d}}}, 
\end{equation}
which is independent of the group index $n_g$. Now let 
\begin{equation}
\mathbf{Z}_{0} \triangleq \begin{bmatrix}
Z_\text{self} &  Z_m \\
Z_m &  Z_\text{self}
\end{bmatrix},
\end{equation}
with $\text{Re}\{Z_\text{self}\}  = R_\text{self}$ and $\text{Re}\{Z_m\}  = R_m$. The input impedance matrix $\mathbf{Z}_{0}$ is computed by the induced EMF method for two side-by-side dipoles~\cite[Ch. 8]{balanis_book}. Since $\mathbf{Z} = \mathbf{I}_{N_g}\otimes \mathbf{Z}_0$, it is straightforward to prove that $\mathbf{Z}_a = \text{blkdiag}(\mathbf{Z}_{0,a} ,\dots, \mathbf{Z}_{0,a})$ using the relationship $\mathbf{Z}_a\mathbf{i}^{\text{opt}}=\mathbf{Z}\mathbf{i}^{\text{opt}}$, where
\begin{align}
&\mathbf{Z}_{0,a} = \begin{bmatrix}
Z_\text{self} + Z_m\frac{I_{n_g+1}(0)}{I_{n_g}(0)} & 0 \\
0 &  Z_\text{self} + Z_m\frac{I_{n_g}(0)}{I_{n_g+1}(0)}
\end{bmatrix} \nonumber \\
&=
 \begin{bmatrix}
Z_\text{self} + Z_m\frac{-R_{m} + R_{\text{self}}e^{-j\kappa \bar{d}}}{R_{\text{self}} -  R_m e^{-j\kappa \bar{d}}} & 0 \\
0 &  Z_\text{self} + Z_m\frac{R_{\text{self}} -  R_m e^{-j\kappa \bar{d}}}{-R_{m} + R_{\text{self}}e^{-j\kappa \bar{d}}}
\end{bmatrix}.
\end{align} 
Therefore, all dipole pairs share a common active impedance matrix whose entries are $[\mathbf{Z}_{0,a}]_{1,1} = 21.84 + j32.89 \ \Omega$, and $[\mathbf{Z}_{0,a}]_{2,2} = 40.87 + j83.89 \ \Omega$, for half-wavelength dipoles made of copper and inter-element distance $\bar{d} = \lambda/5$. This is a unique feature of the proposed \ac{NULA}/\ac{NUPA}. In contrast, a superdirective \ac{ULA}/\ac{UPA} would need a different matching impedance for each port because the current amplitude is not uniform along the ports (akin to the source voltages). Thus, the derived architecture simplifies also single-port matching. Finally, compared to an uncoupled \ac{ULA}/\ac{UPA} with $Z_\text{self} = 75.94 + j41.76$~$\Omega$, the active impedances $[\mathbf{Z}_{0,a}]_{1,1} $ and $[\mathbf{Z}_{0,a}]_{2,2} $  are not very large, and thus could be easily matched.

\subsection{Multiuser Transmissions with Imperfect CSI}
We now consider that the \ac{BS} simultaneously transmits to $K$ users using \ac{MRT}. The users' directions are not fixed, and thus the \ac{BS} acquires the channel estimates $\{\hat{\mathbf{h}}_k\}_{k=1}^K$ through the \ac{OMP} estimator of Section~\ref{sec:omp}. In the \ac{UPA} case, the dictionary is $\bar{\mathbf{H}} = \left[\mathbf{a}\left(\bar{\theta}_0,\bar{\phi}_0\right), \dots, \mathbf{a}\left(\bar{\theta}_{G_z-1},\bar{\phi}_{G_x-1}\right) \right]$. The \ac{BS} treats those estimates as the true channels in the beamforming stage, i.e., $\mathbf{w}_k = \hat{\mathbf{h}}^*_k/\|\hat{\mathbf{h}}_k\|$. The achievable rate of user $k$ is then specified as $R_k = B\log_2(1 + \text{SINR}_k)$, where $\text{SINR}_k$ is given by~\eqref{eq:sinr}. Note that this rate is achieved under the assumption that user $k$ knows $|\mathbf{h}_k^T \mathbf{w}_k|^2$ and $\sum_{i\neq k}  |\mathbf{h}_k^T \mathbf{w}_i|^2$ in the decoding stage. These are scalars and, hence, are easy to be estimated. For the performance evaluation, the primary metrics are the sum-rate, minimum rate, and total \ac{EE} defined as $\sum_{k=1}^KR_k$, $\min_k\{R_k\}$, and $\sum_{k=1}^KR_k/P_c$, respectively. 

Figure~\ref{fig:Fig7} shows the results for a two-user transmission. As observed, the \ac{NUPA} boosts the mean \ac{EE} by $29\%$ without compromising the sum or minimum data rate. Regarding the \ac{OMP} estimator, we calculate the \ac{NSE} defined as $\text{NSE}\triangleq \frac{1}{K}\sum_{k=1}^K\left\|\mathbf{h}_k-\hat{\mathbf{h}}_k\right\|^2 \big/ \|\mathbf{h}_k\|^2$.
The training overhead per user is $N_{\text{beam}}=0.8N\approx 564$ pilot beams for the \ac{NUPA}, whilst $N_{\text{beam}}\approx 820$ beams for the \ac{UPA}. Importantly, Fig.~\ref{fig:Fig8} indicates that channel estimation accuracy is similar for both arrays, although the \ac{NUPA} employs 256 pilots less. Consequently, it has the potential to reduce also the \ac{CSI} acquisition overhead of massive \ac{MIMO} \ac{BS}.
\begin{figure*}[t]
	\centering
	\begin{subfigure}{0.5\textwidth}
		\centering
		\includegraphics[width=0.94\linewidth]{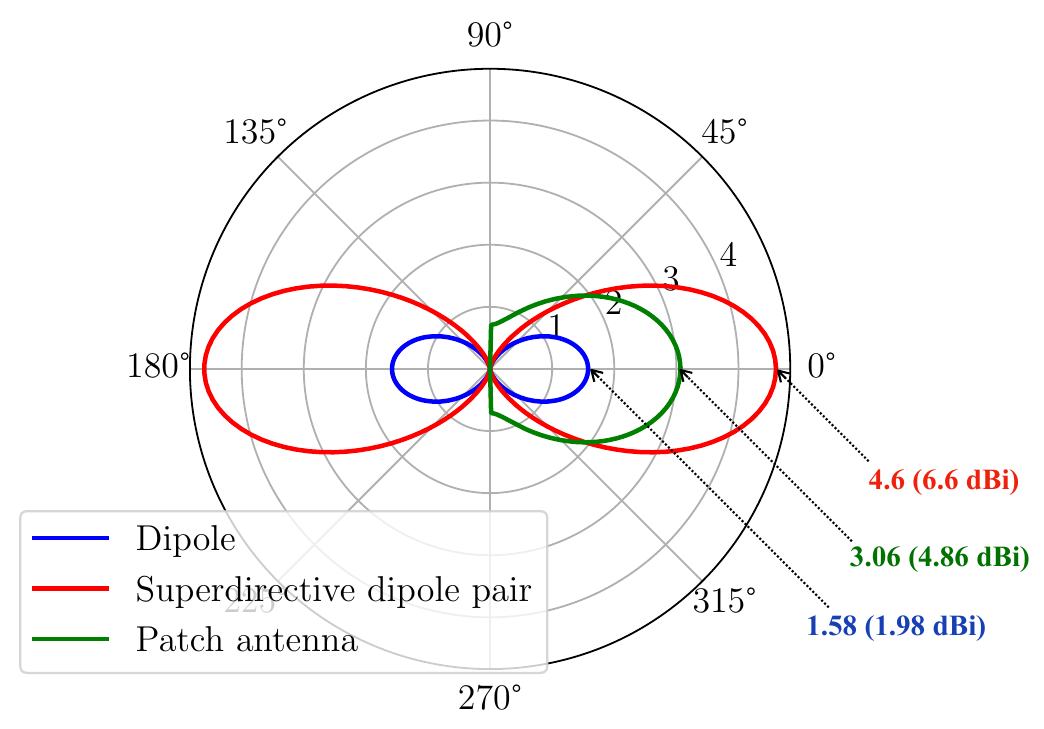}
		\subcaption{}
	\end{subfigure}~
	\begin{subfigure}{0.5\textwidth}
		\centering
		\includegraphics[width=0.6\linewidth]{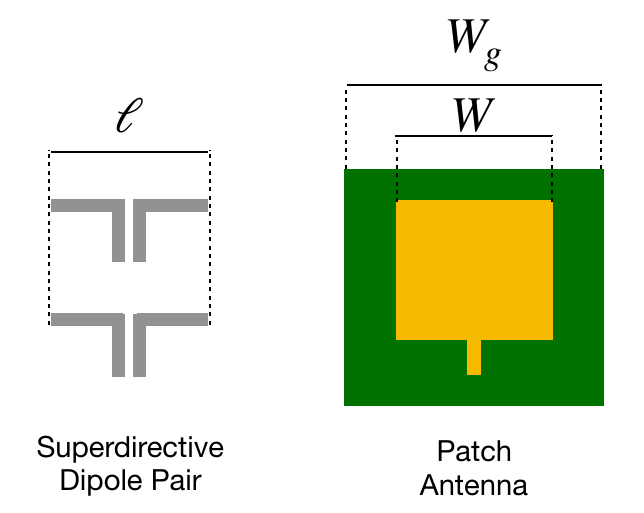}
		\subcaption{}
	\end{subfigure}
	\caption{(a) Gain patterns at the vertical plane $\phi=0$. The spacing between dipoles is $\bar{d}=\lambda/5$. (b) Sketch of patch antenna and superdirective pair geometries.}
	\label{fig:Fig9}
\end{figure*}
\subsection{Comparison with Patch Antennas}
Microstrip antennas, also known as patch antennas, are widely adopted in real-world \ac{mmWave} and \ac{THz} MIMO systems due to their planar geometry, directional radiation characteristics, and easy fabrication into printed circuit boards~\cite{thz_antenna_overview}. As such, it is of practical interest to investigate how our superdirective design compares with a typical patch antenna array. Here, we would like to stress that the proposed superdirective pair can be readily realized with printed dipoles in order to have a planar radiating structure~\cite{printed_dipole_array}. In the sequel, we consider a copper patch of length $L$ and width $W$ on top of a grounded dielectric substrate of thickness $h$, dielectric constant $\epsilon_r$ and loss tangent $\tan\delta$. To ensure high radiation efficiency, the substrate thickness is selected in the range $ 0.025\lambda\leq h \leq 0.05\lambda$, where $\lambda$ is the free-space wavelength. Since the fields generated by the antenna propagate in two different media, a homogeneous medium is assumed with effective dielectric constant~\cite{balanis_book} 
\begin{equation}
\epsilon_{\text{reff}} = \frac{\epsilon_r+1}{2} + \frac{\epsilon_r-1}{2}\frac{1}{\sqrt{1+12h/W}}.
\end{equation}

\subsubsection{Antenna Gain}
For the dominant transverse magnetic $\text{TM}_{010}$ mode, the width and length of the copper patch are chosen as~\cite{balanis_book} 
\begin{align}
W = \frac{c}{2 f_r}\sqrt{\frac{2}{\epsilon_r+1}}, 
\end{align}
and
\begin{align}
L = \frac{c}{2f_r\sqrt{\epsilon_{\text{reff}}}} - 2\Delta L,
\end{align}
where $f_r$ denotes the resonant frequency of the antenna, and 
\begin{equation}
\Delta L = 0,412 h \frac{(\epsilon_{\text{reff}} +0.3)(W/h + 0.264)}{(\epsilon_{\text{reff}} -0.258)(W/h + 0.8)}.
\end{equation}
Using these equations, the dimensions of the radiator are calculated for given $f_r, h, \epsilon_r$, and $\tan\delta$. Next, the azimuth $E_{\theta}$ and polar $E_{\phi}$ components of the electric field of the antenna are specified by~\cite[Ch. 4]{microstrip_book}
\begin{align}\label{eq: e_field_patch1}
E_{\theta} &= \frac{2h}{\pi}\frac{\sin\left(\frac{\pi W}{\lambda}\sin\theta\sin\phi\right)}{\sin\theta\sin\phi}\cos\left(\frac{\pi  L_{\text{reff}}}{\lambda}\sin\theta\cos\phi\right)\cos\phi, \\
E_{\phi} & = -\frac{2h}{\pi}\frac{\sin\left(\frac{\pi W}{\lambda}\sin\theta\sin\phi\right)}{\sin\theta\sin\phi} \nonumber \\
&\times \cos\left(\frac{\pi  L_{\text{reff}}}{\lambda}\sin\theta\cos\phi\right)\cos\theta\sin\phi,\label{eq: e_field_patch2}
\end{align}
where $L_{\text{reff}} \triangleq L +\Delta L$ is the effective length. Note that \eqref{eq: e_field_patch1}-\eqref{eq: e_field_patch2} are valid for polar angles $0\leq \theta \leq \pi/2$ due to the presence of the ground plane. The directivity of the antenna is defined as $D(\theta,\phi)\triangleq 4\pi U/P_{\text{rad}}$, where $U = \frac{1}{2\eta}(E^2_{\theta} +  E^2_{\phi})$ is the radiation intensity and $P_{\text{rad}}  = \int_0^{2\pi}\!  \int_0^{\pi/2} U \sin\theta \text{d}\theta \text{d}\phi  $ is the radiated power.  

For sufficiently thin substrate, the surface wave losses can be neglected, and hence the radiation efficiency is determined by
\begin{equation}
\eta_{\text{rad}} = \frac{P_{\text{rad}}}{P_{\text{rad}} + P_c + P_d},
\end{equation}
where $P_c$ and $P_d$ denote the power dissipated in the conducting patch and dielectric substrate, respectively. These are given by the known closed-form expressions~\cite[Ch. 4]{microstrip_book}
\begin{align}
P_d &=\frac{1}{4}\omega \epsilon_0\epsilon_rhLW\tan\delta, \\
P_c &=\frac{\omega \epsilon_0\epsilon_rLW}{4\sigma_c R_s},
\end{align}
where $\epsilon_0$ is the permittivity of free space, $\omega$ is the angular frequency, and $R_s = \sqrt{\omega \mu/(2\sigma_c)}$ is the sheet resistance of the copper patch. Finally, the element gain is $G_e(\theta,\phi) = \eta_{\text{rad}} D(\theta,\phi)$.

Figure \ref{fig:Fig9}(a) shows the gain pattern of a single half-wavelength dipole, a supedirective dipole pair, and a rectangular patch antenna with copper and a Polydimethylsiloxane  substrate ($\epsilon_r = 2.35, \tan\delta = 0.03$) of thickness $h = 0.03\lambda$~\cite{patch_antenna}. The estimated patch antenna directivity at broadside is $D(0,0) = 7.1$ dBi with radiation efficiency $\eta_{\text{rad}} = 60\%$, yielding a maxum gain $G_e(0,0) = 4.86$ dBi in perfect agreement with~\cite{patch_antenna, patch_antenna_array}. On the other hand, the superdirective dipole pair offers a $6.6$ dBi gain, and exhibits sufficient directionality at the element level. 
Regarding the footprint of the radiators, the width of the patch antenna is determined by the width of the grounded substrate, as depicted in Fig. \ref{fig:Fig9}(b), and is given by $W_g = 6h + W \approx 0.56\lambda$~\cite{microstrip_book}. Therefore, it is slightly larger than that of the half-wavelength dipoles comprising the superdirective pair.
\begin{figure}[t]
	\centering
	\includegraphics[width=0.88\linewidth]{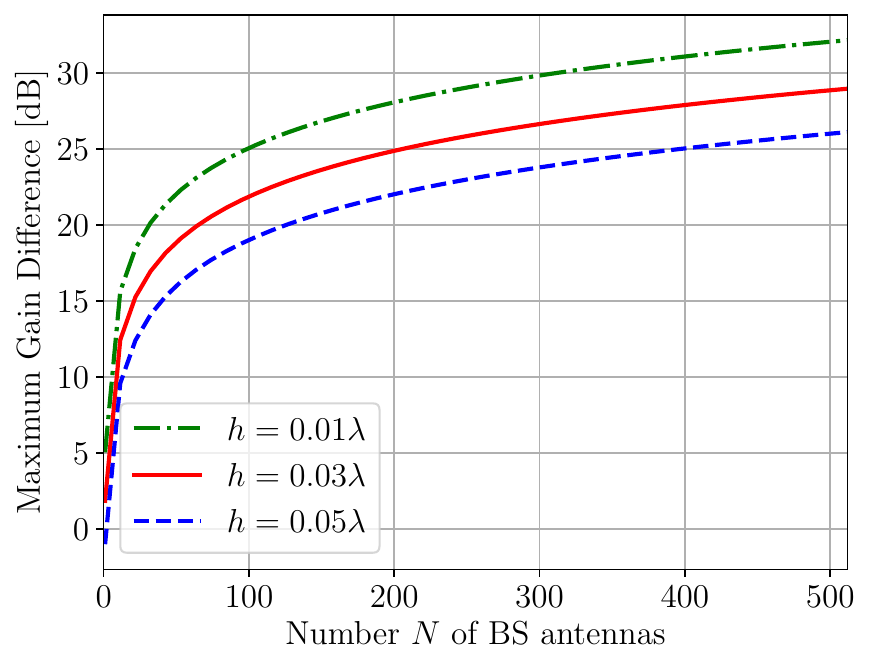}
	\caption{Broadside gain difference versus the number of BS antennas for various substrate heights.}
	\label{fig:Fig10}
\end{figure}

\subsubsection{Array Gain}
We now consider a linear array of $N$ elements. Both the superdirective pair and the patch antenna have nulls in the radiation pattern at $\theta = \pi/2$, and hence the elements can be placed in a linear configuration with subwavelength spacing and negligible mutual coupling. Therefore, the maximum broadside array gain is $G(0,0) = G_e(0,0) N$. Figure~\ref{fig:Fig10} shows the broadside gain difference between the patch and superdirective antenna arrays. As seen for $h=0.03\lambda$, the performance improvement is more than $20$ dB for $N>100$ antennas. More importantly, this is attained with a compact configuration as the patch antenna array will be $N(W_g - \lambda/2) > 6\lambda$ wider for $N> 100$. Consequently, the proposed grouping architecture can enable the realization of low-profile MIMO transceivers. 

\section{Conclusions}\label{sec:conclusions}
In this paper, we introduced the novel concept of superdirective dipole pairs. Specifically, we first derived a comprehensive array model that captures the physics of mutual coupling and ohmic losses of nonideal antennas. Capitalizing on the derived model, we then studied the implementation aspects of superdirectivity, namely the impedance matching and hybrid beamforming problems. To surmount the challenges of superdirective \ac{ULA}s/\ac{UPA}s, we proposed to partition the \ac{BS} array into multiple two-element groups of sub-wavelength spacing. The resulting \ac{NULA}/\ac{NUPA} facilitates multi-port impedance matching, which is optimal for any beamforming angle. More importantly, it enables the realization of superdirectivity with a single \ac{RF} chain per beam. Afterwards, we pursued a performance analysis in terms of achievable rate and \ac{EE} under perfect and imperfect \ac{CSI}. To this end, approximate closed-form expressions for the signal and multi-user interference powers were provided under \ac{MRT}. The channel estimation problem was also addressed by employing \ac{OMP} along with a coupling-aware dictionary. Numerical results were finally provided demonstrating that the proposed method boosts the~\ac{EE} of \ac{THz} massive \ac{MIMO} without compromising the data transmission and channel estimation performances. As a result, arrays of superdirective dipole pairs can be a promising approach for realizing energy-efficient \ac{MIMO} antennas with sharp beamforming capabilities. 

\section*{Acknowledgements}
This project has received funding from the European Research Council (ERC) under the European Union’s Horizon 2020 research and innovation programme (grant agreement No. 101001331). The work of H. Q. Ngo was supported by the U.K. Research and Innovation Future Leaders Fellowships under Grant MR/X010635/1.

\section*{Appendix A}
Based on the radiation equations,  the electric field is specified as~\cite[Ch. 3]{balanis_book}
\begin{align}
\mathbf{E}(r,\theta,\phi) = -j\eta \frac{k e^{-j \kappa r}}{4\pi r } (A_{\theta}\mathbf{e}_{\theta} +  A_{\phi}\mathbf{e}_{\phi}),
\end{align}
where 
\begin{align}
A_\theta &=\int_{-\ell/2}^{\ell/2} I(x')\cos\theta\cos\phi e^{j\kappa x'\cos\phi\sin\theta}\text{d}x  \nonumber\\
&= \frac{I(0) \cos\theta\cos\phi}{\sin(\kappa\ell/2)} \!\!\int_{-\ell/2}^{\ell/2}  \sin\left(\kappa\ell/2-\kappa|x'|\right)e^{j\kappa x'\cos\phi\sin\theta}\text{d}x ,
\end{align}
and
\begin{align}
A_\phi &=\int_{-\ell/2}^{\ell/2} -I(x')\sin\phi e^{j\kappa x'\cos\phi\sin\theta}\text{d}x \nonumber \\
&=- \frac{I(0)\sin\phi }{\sin(\kappa\ell/2)} \int_{-\ell/2}^{\ell/2}  \sin\left(\kappa\ell/2-\kappa|x'|\right)e^{j\kappa x'\cos\phi\sin\theta}\text{d}x.
\end{align}
Utilizing the identity~\cite{balanis_book} 
\begin{equation}
\int e^{a x}\sin(\beta x + \gamma)\text{d}x  = \frac{e^{ax}}{a^2 + \beta^2}[a\sin(\beta x +\gamma) - \beta \cos(\beta x + \gamma)],
\end{equation}
for $a = j\kappa\cos\phi\sin\theta$, $\beta = \kappa$, and $\gamma = \kappa\ell/2$, and after some algebraic manipulations, we get 
\begin{align}
&\int_{-\ell/2}^{\ell/2}  \sin\left(\kappa\ell/2-\kappa|x'|\right)e^{j\kappa x'\cos\phi\sin\theta}\text{d}x = \nonumber \\
&
=\frac{2}{\kappa}\frac{\cos(\kappa\ell/2\cos\phi\sin\theta) - \cos(\kappa\ell/2)}{\sin^2\phi + \cos^2\phi\cos^2\theta}.
\end{align}
Combining those equations yields the field expression in~\eqref{eq:e_field}. 

\section*{Appendix B}
The inverse matrix of $\mathbf{Z}_0$ is
  \begin{equation}\label{eq:inverse_Z0}
\text{Re}\{\mathbf{Z}_0\}^{-1}  = \frac{1}{R_{\text{self}}^2  - R^2_m}\begin{bmatrix}
R_{\text{self}}   & -R_m \\
-R_m & R_{\text{self}} 
\end{bmatrix}.
\end{equation}
We now need to calculate the square root of the $2\times 2$ matrix $\text{Re}\{\mathbf{Z}_0\}^{-1}$. To do so, we utilize the lemma~\cite{square_root_matrix}
\begin{equation}\label{eq:lemma_square_root}
\mathbf{A}^{1/2} =\frac{1}{t} \begin{bmatrix}
a_{11} +  s & a_{12} \\
a_{21} & a_{22} + s  
\end{bmatrix},
\end{equation}
for
\begin{equation}
\mathbf{A} = \begin{bmatrix}
a_{11}  & a_{12} \\
a_{21} & a_{22}  
\end{bmatrix},
\end{equation}
where $s = \sqrt{a_{11}a_{22} - a_{12}a_{21}}$ and $t=\sqrt{a_{11} + a_{22} + 2s}$. In our case, we have that 
\begin{align}
s = \frac{1}{\sqrt{R_{\text{self}}^2 - R_m^2}}, \quad t & = \sqrt{\frac{2R_{\text{self}}  + 2\sqrt{R_{\text{self}} ^2 - R_m^2}}{R_{\text{self}} ^2 - R_m^2}}.
\end{align}
Applying~\eqref{eq:lemma_square_root} to $\text{Re}\{\mathbf{Z}_0\}^{-1}$, and after basic algebra, yields
\begin{align}
\text{Re}\{\mathbf{Z}_0\}^{-1/2} &= \frac{1}{\sqrt{2R_{\text{self}}  + 2\sqrt{R_{\text{self}} ^2 - R_m^2}}} \nonumber \\
&\times \begin{bmatrix}
\frac{R_{\text{self}} }{\sqrt{R_{\text{self}} ^2 - R_m^2}} + 1& - \frac{R_m}{\sqrt{R_{\text{self}} ^2 - R_m^2}} \\
- \frac{R_m}{\sqrt{R_{\text{self}} ^2 - R_m^2}}  & \frac{R_{\text{self}} }{\sqrt{R_{\text{self}} ^2 - R_m^2}} + 1
\end{bmatrix},
\end{align}
and 
\begin{align}
\text{Re}&\{\mathbf{Z}_0\}^{-1/2}\mathbf{a}_0(\theta) = \frac{1}{\sqrt{2R_{\text{self}}  + 2\sqrt{R_{\text{self}} ^2 - R_m^2}}}  \nonumber \\
&\times
\begin{bmatrix}
\frac{R_{\text{self}} }{\sqrt{R_{\text{self}}^2 - R_m^2}} + 1 - \frac{R_m}{\sqrt{R_{\text{self}}^2 - R_m^2}}e^{-j\kappa \bar{d}\cos\theta} \\
- \frac{R_m}{\sqrt{R_{\text{self}}^2 - R_m^2}}  + \left(\frac{R_{\text{self}}}{\sqrt{R_{\text{self}}^2 - R_m^2}} + 1\right)e^{-j\kappa \bar{d}\cos\theta} 
\end{bmatrix}.
\end{align}
Lastly, calculating the magnitudes of the entries of $\text{Re}\{\mathbf{Z}_0\}^{-1/2} \mathbf{a}_0(\theta)$, and  some algebraic manipulations, we obtain~\eqref{eq:mag_pair}. Also, because $\mathbf{Z}\approx \mathbf{I}_{N_g}\otimes \mathbf{Z}_0$ in the proposed \ac{NULA},~\eqref{eq:optimal_matching} becomes
\begin{align}\label{eq:Z_M_nula}
&\mathbf{Z}_M \!= \! \begin{bmatrix}
-j\text{Im}\{Z_s\} \mathbf{I}_{N_g}\otimes\mathbf{I}_2  & -j\sqrt{R_s}\mathbf{I}_{N_g}\otimes\text{Re}\{\mathbf{Z}_0\}^{1/2} \\ 
-j\sqrt{R_s}\mathbf{I}_{N_g}\otimes\text{Re}\{\mathbf{Z}_0\}^{1/2} & -j\mathbf{I}_{N_g}\otimes\text{Im}\{\mathbf{Z}_0\}
\end{bmatrix}.
\end{align}

We now partition the vectors of voltages and currents at the input and output of the impedance matching network as $\mathbf{v}_M = [\mathbf{v}_{M,0},\dots, \mathbf{v}_{M,N_g-1}]^T$, 
$\mathbf{i}_M =[\mathbf{i}_{M,0},\dots, \mathbf{i}_{M,N_g-1}]^T$, 
$\mathbf{v} =[\mathbf{v}_{0},\dots, \mathbf{v}_{N_g-1}]^T$, and  $\mathbf{i} =[\mathbf{i}_{0},\dots, \mathbf{i}_{N_g-1}]^T$, where $\mathbf{v}_{M,n_g}=[v_{M,n_g},  v_{M,n_g+1}]^T$ is the vector of voltages at the $n_g$th input port pair; $\mathbf{i}_{M,n_g}$, $\mathbf{v}_{n_g}$, and $\mathbf{i}_{n_g}$ are defined similarly. By using~\eqref{eq:Z_M_nula} and the previous decompositions,~\eqref{eq:matching_network} is recast as
\begin{align}
&\begin{bmatrix}
\begin{bmatrix}
\mathbf{v}_{M,0}\\
\mathbf{v}_{0}
\end{bmatrix}
&\cdots &
\begin{bmatrix}
\mathbf{v}_{M,N_g-1}\\
\mathbf{v}_{N_g-1}
\end{bmatrix}
\end{bmatrix}^T = \nonumber \\
&= \begin{bmatrix}
\mathbf{Z}_{M,0}
\begin{bmatrix}
\mathbf{i}_{M,0}\\
-\mathbf{i}_{0}
\end{bmatrix}
&
\cdots 
&\mathbf{Z}_{M,0}
\begin{bmatrix}
\mathbf{i}_{M,N_g-1}\\
-\mathbf{i}_{N_g-1}
\end{bmatrix}
\end{bmatrix}^T,
\end{align}
which shows that $\mathbf{Z}_M$ comprises $N_g$ four-port matching networks, which are independent of each other.

\section*{Appendix C}
We have that 
\begin{align}
&\frac{\left|\mathbf{a}^H(\theta_k)\text{Re}\left\{\bar{\mathbf{Z}}_{\text{approx}}\right\}^{-1}\mathbf{a}(\theta_i)\right|^2}{\mathbf{a}^H(\theta_i)\text{Re}\left\{\bar{\mathbf{Z}}_{\text{approx}}\right\}^{-1}\mathbf{a}(\theta_i)} \nonumber \\
&=   \frac{\left|(\mathbf{a}^H_g(\theta_k)\otimes \mathbf{a}^H_0(\theta_k))\!\left(\mathbf{I}_{N_g}\otimes \text{Re}\left\{\bar{\mathbf{Z}}_0\right\}^{-1}\right)\!(\mathbf{a}_g(\theta_i)\otimes \mathbf{a}_0(\theta_i))\right|^2}{(\mathbf{a}^H_g(\theta_i)\otimes \mathbf{a}^H_0(\theta_i))\!\left(\mathbf{I}_{N_g}\otimes \text{Re}\left\{\bar{\mathbf{Z}}_0\right\}^{-1}\right)\!(\mathbf{a}_g(\theta_i)\otimes \mathbf{a}_0(\theta_i))} \nonumber \\
& = \!\frac{\left|\mathbf{a}^H_0(\theta_k)\text{Re}\left\{\bar{\mathbf{Z}}_0\right\}^{-1}\mathbf{a}_0(\theta_i)\right|^2}{ \mathbf{a}^H_0(\theta_i)\text{Re}\left\{\bar{\mathbf{Z}}_0\right\}^{-1}\mathbf{a}_0(\theta_i)}\frac{ \left|\mathbf{a}^H_g(\theta_k)\mathbf{a}_g(\theta_i)\right|^2}{N_g},
\end{align}
where the identity $(\mathbf{A}\otimes \mathbf{B})(\mathbf{C}\otimes \mathbf{D}) = (\mathbf{A} \mathbf{C})\otimes(\mathbf{B}\mathbf{D})$ has been applied twice. Lastly, 
\begin{align}
&\frac{ \left|\mathbf{a}^H_g(\theta_k)\mathbf{a}_g(\theta_i)\right|^2}{N_g} 
= \frac{1}{N_g}\left| \sum_{n_g=0}^{N_g-1}e^{-j\kappa n_g(d_g + (\bar{N}-1)\bar{d})(\cos\theta_k-\cos\theta_i)}\right |^2  \nonumber \\
& =  \frac{1}{N_g}\left|\frac{1 - e^{-j  \kappa N_g(d_g + (\bar{N}-1)\bar{d})(\cos\theta_k-\cos\theta_i)}}{1 - e^{-j\kappa(d_g + (\bar{N}-1)\bar{d})(\cos\theta_k-\cos\theta_i)}}\right |^2\\
& = N_g|D_{N_g}\left(\kappa(d_g + (\bar{N}-1)\bar{d})(\cos\theta_k-\cos\theta_i)\right)|^2,
\end{align}
which completes the proof.

\begin{IEEEbiography}[{\includegraphics[width=1in,height=1.25in,clip,keepaspectratio]{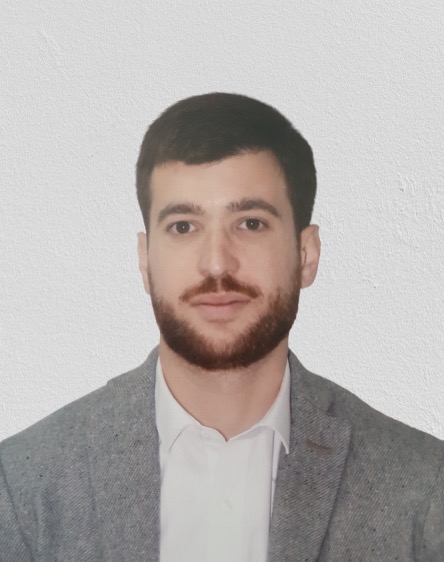}}]
	{Konstantinos Dovelos} received the Diploma (M.Eng.) degree in electrical and computer engineering from the Aristotle University of Thessaloniki, Greece, in 2016,  and the Ph.D. degree from Universitat Pompeu Fabra, Spain, in 2021. From September 2021 through September 2022, he was with the Centre for Wireless Innovation (CWI) at Queen's University Belfast, U.K., working as a Postdoctoral Research Fellow. He is currently an R\&D Enginner at Meta Materials Inc., Greece. His research interests span massive MIMO architectures, superdirectivity, antennas, metasurfaces, and electromagnetic information theory. 
\end{IEEEbiography}

\begin{IEEEbiography}[{\includegraphics[width=1in,height=1.25in,clip,keepaspectratio]{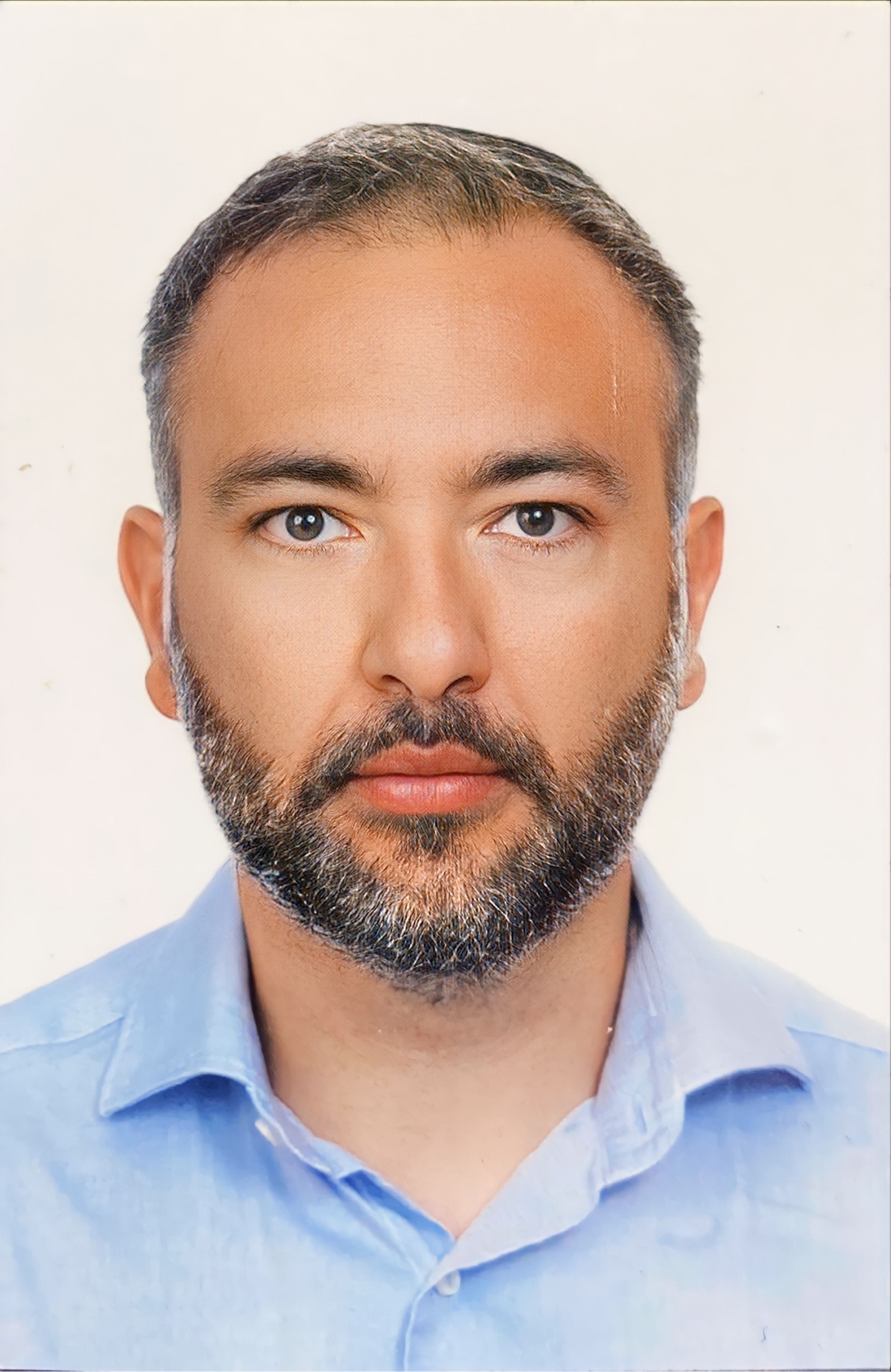}}]
{Stylianos D. Assimonis} received his Diploma (5~years) and Ph.D. degrees in Electrical and Computer Engineering from Aristotle University of Thessaloniki, Greece. Currently, he holds the position of Lecturer (Assistant Professor) at the School of Electronics, Electrical Engineering and Computer Science in Queen's University, Belfast, UK. His primary research interests encompass Electromagnetic Periodic Structures (including metasurfaces and reconfigurable intelligent surfaces), RF Engineering (wireless sensing, Internet of Things (IoT), and RF Energy Harvesting), and Antennas (spanning super-directive antennas, electrically small antennas, millimeter-wave (mm-Wave) antennas, and electronically steerable parasitic array radiator (ESPAR) antennas).

Dr. Assimonis was honored with the Post-Doctoral Scholarship for Excellence by the Research Committee of Aristotle University of Thessaloniki and by the Centre for Wireless Innovation (CWI) at Queen's University Belfast, UK, in 2012 and 2016, respectively. He has co-authored numerous research papers in the fields of Electromagnetics and RF Engineering, some of which have received distinguished paper awards at prominent events such as Metamaterials 2013, the 2014 IEEE RFID-TA, and the 2015 5th COST IC1301 Workshop. He currently serves as an Editor for Nature Scientific Reports and the MDPI Journal of Low Power Electronics and Applications. In 2019 and 2021, he acted as a Guest Editor for the MDPI Journal of Low Power Electronics and Applications and the MDPI Sensors, respectively.
\end{IEEEbiography}

\begin{IEEEbiography}[{\includegraphics[width=1in,height=1.25in,clip,keepaspectratio]{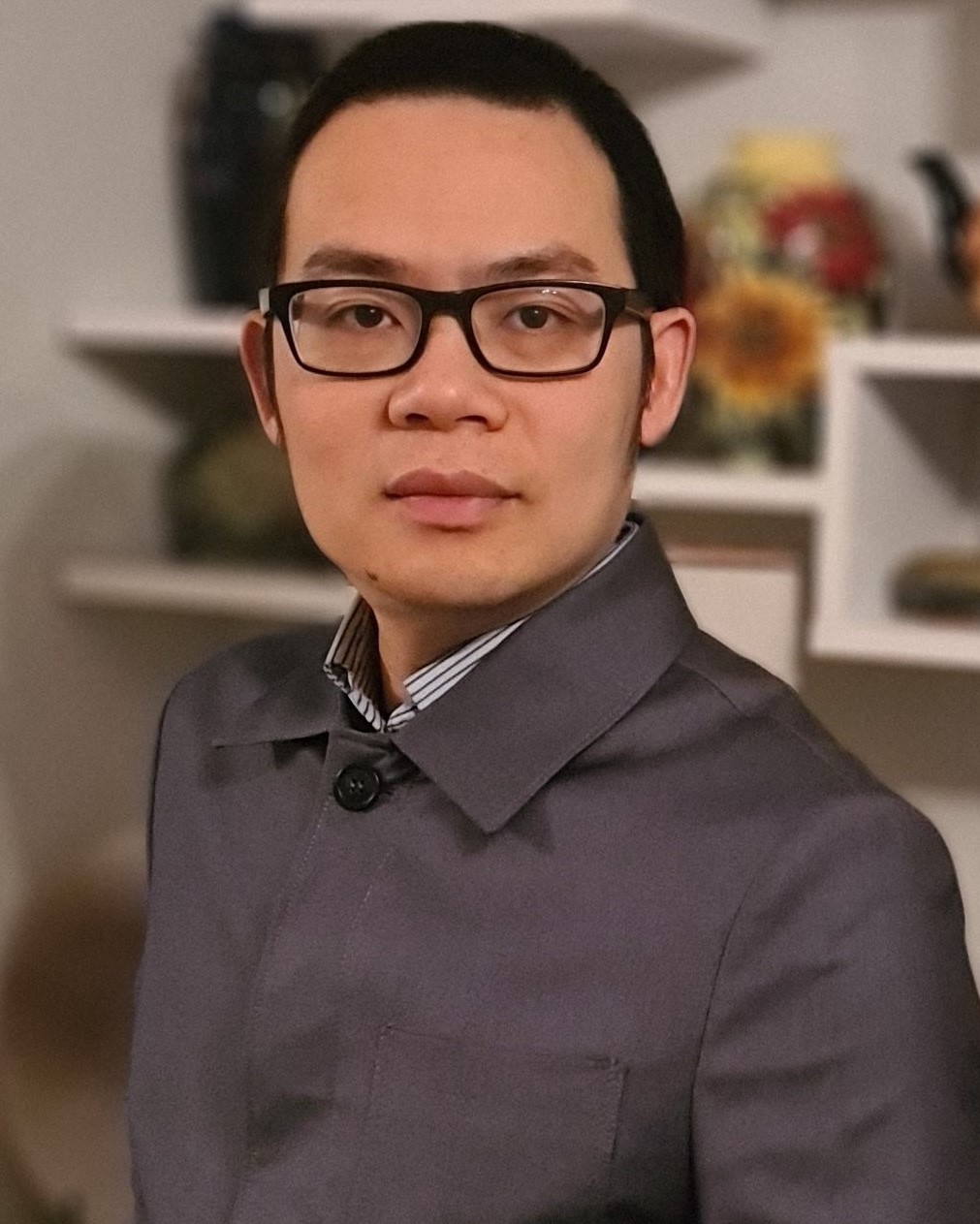}}]
{Hien Quoc Ngo} is currently a Reader with Queen's University Belfast, U.K. His main research interests include massive MIMO systems, cell-free massive MIMO, reconfigurable intelligent surfaces, physical layer security, and cooperative communications. He has co-authored many research papers in wireless communications and co-authored the Cambridge University Press textbook \emph{Fundamentals of Massive MIMO} (2016).

He received the IEEE ComSoc Stephen O. Rice Prize in 2015, the IEEE ComSoc Leonard G. Abraham Prize in 2017, and the Best Ph.D. Award from EURASIP in 2018. He also received the IEEE Sweden VT-COM-IT Joint Chapter Best Student Journal Paper Award in 2015. He was awarded the UKRI Future Leaders Fellowship in 2019. He serves as the Editor for the IEEE Transactions on Wireless Communications, IEEE Transactions on Communications, the Digital Signal Processing, and the Physical Communication (Elsevier). He was a Guest Editor of IET Communications, and a Guest Editor of IEEE ACCESS in~2017.
\end{IEEEbiography}

\begin{IEEEbiography}[{\includegraphics[width=1in,height=1.25in,clip,keepaspectratio]{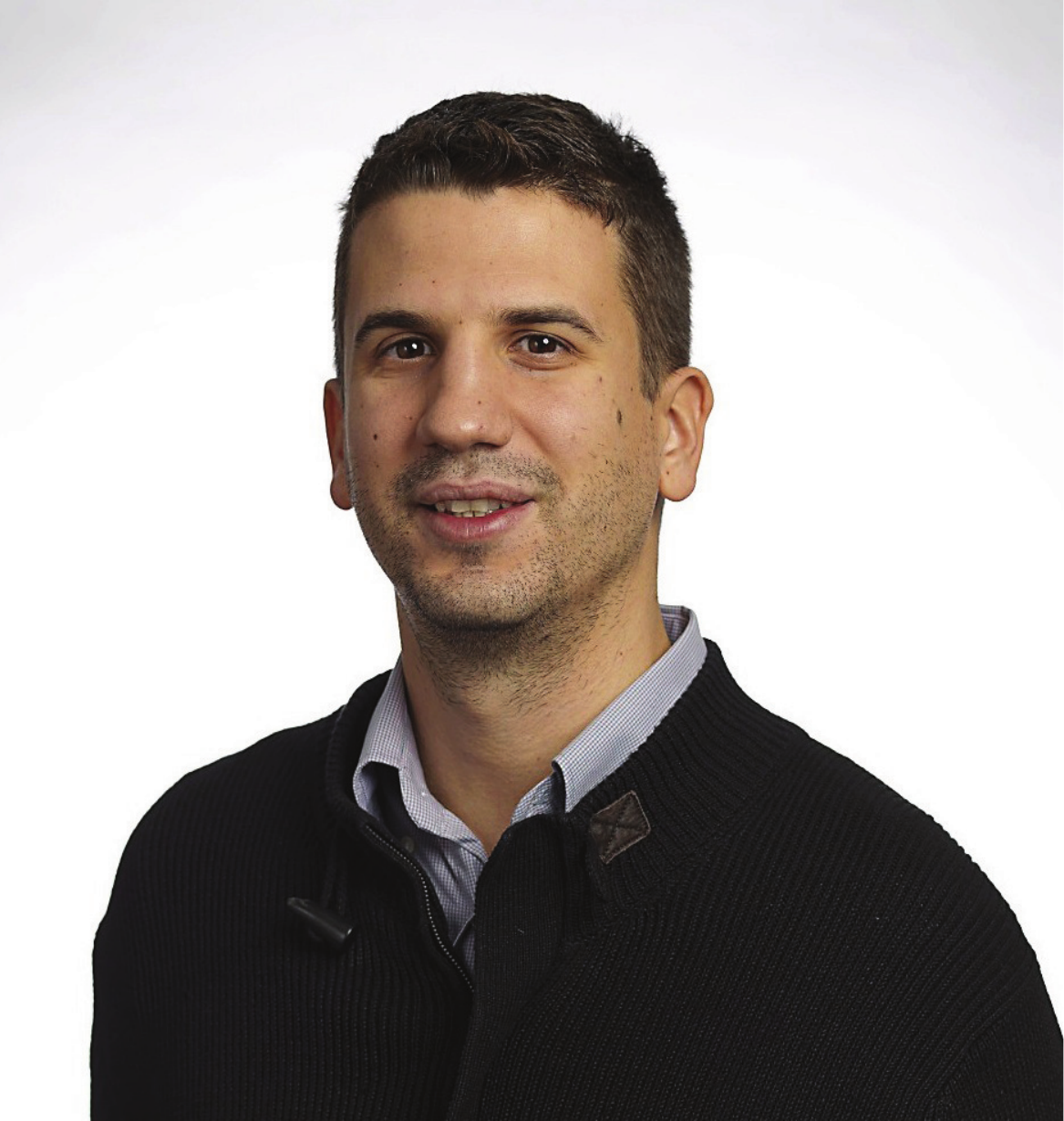}}]
{Michail Matthaiou}(Fellow, IEEE) was born in Thessaloniki, Greece in 1981. He obtained the Diploma degree (5 years) in Electrical and Computer Engineering from the Aristotle University of Thessaloniki, Greece in 2004. He then received the M.Sc. (with distinction) in Communication Systems and Signal Processing from the University of Bristol, U.K. and Ph.D. degrees from the University of Edinburgh, U.K. in 2005 and 2008, respectively. From September 2008 through May 2010, he was with the Institute for Circuit Theory and Signal Processing, Munich University of Technology (TUM), Germany working as a Postdoctoral Research Associate. He is currently a Professor of Communications Engineering and Signal Processing and Deputy Director of the Centre for Wireless Innovation (CWI) at Queen’s University Belfast, U.K. after holding an Assistant Professor position at Chalmers University of Technology, Sweden. His research interests span signal processing for wireless communications, beyond massive MIMO, intelligent reflecting surfaces, mm-wave/THz systems and deep learning for communications.
	
Dr. Matthaiou and his coauthors received the IEEE Communications Society (ComSoc) Leonard G. Abraham Prize in 2017. He currently holds the ERC
Consolidator Grant BEATRICE (2021-2026) focused on the interface between information and electromagnetic theories. He was awarded the prestigious 2018/2019 Royal Academy of Engineering/The Leverhulme Trust Senior Research Fellowship and also received the 2019 EURASIP Early Career Award. His team was also the Grand Winner of the 2019 Mobile World Congress Challenge. He was the recipient of the 2011 IEEE ComSoc Best Young Researcher Award for the Europe, Middle East and Africa Region and a co-recipient of the 2006 IEEE Communications Chapter Project Prize for the best M.Sc. dissertation in the area of communications. He has co-authored papers that received best paper awards at the 2018 IEEE WCSP and 2014 IEEE ICC. In 2014, he received the Research Fund for International Young Scientists from the National Natural Science Foundation of China. He is currently the Editor-in-Chief of Elsevier Physical Communication, a Senior Editor for \textsc{IEEE Wireless Communications Letters} and \textsc{IEEE Signal Processing Magazine}, and an Associate Editor for \textsc{IEEE Transactions on Communications}. He is an IEEE Fellow.
\end{IEEEbiography}

\end{document}